\documentclass{jfm}

\usepackage{natbib}
\usepackage{amsmath,amssymb}

\usepackage{color}
\usepackage{graphicx}
\usepackage{subfigure}

\ifCUPmtlplainloaded \else
  \checkfont{eurm10}
  \iffontfound
    \IfFileExists{upmath.sty}
      {\typeout{^^JFound AMS Euler Roman fonts on the system,
                   using the 'upmath' package.^^J}%
       \usepackage{upmath}}
      {\typeout{^^JFound AMS Euler Roman fonts on the system, but you
                   dont seem to have the}%
       \typeout{'upmath' package installed. JFM.cls can take advantage
                 of these fonts,^^Jif you use 'upmath' package.^^J}%
      }
  \else
  \fi
\fi

\ifCUPmtlplainloaded \else
  \checkfont{msam10}
  \iffontfound
    \IfFileExists{amssymb.sty}
      {\typeout{^^JFound AMS Symbol fonts on the system, using the
                'amssymb' package.^^J}%
       \usepackage{amssymb}%
       \let\le=\leqslant  
         
      }{}
  \fi
\fi

\ifCUPmtlplainloaded \else
  \IfFileExists{amsbsy.sty}
    {\typeout{^^JFound the 'amsbsy' package on the system, using it.^^J}%
     \usepackage{amsbsy}}
    {}
\fi

\newsavebox{\astrutbox}
\sbox{\astrutbox}{\rule[-5pt]{0pt}{20pt}}

\newcommand\nc{\newcommand}
\nc\beq{\begin{equation}}
\nc\eeq{\end{equation}}
\nc\bea{\begin{eqnarray}}
\nc\eea{\end{eqnarray}}
\nc\bes{\begin{eqnarray*}}
\nc\ees{\end{eqnarray*}}
\nc{\vect}[1]{\mbox{\boldmath $#1$}}
\nc\cB{{\cal{B}}}
\nc\cC{{\cal{C}}}
\nc\cN{{\cal{N}}}
\nc\cG{{\cal{G}}}
\newcommand{\pa}{\partial}
\newcommand{\pad}[2]{\frac{\pa #1}{\pa #2}}

\newcommand{\mylab}[1]{\label{#1}}

\title{Modelling spreading dynamics of liquid crystals in three spatial dimensions}

\author[T.-S. Lin, L. Kondic, U. Thiele and L.J. Cummings]%
{T.\ls-\ls S.\ns L\ls I\ls N\ls$^1$%
,\ns
L.\ns K\ls O\ls N\ls D\ls I\ls C$^2$
,\ns
U.\ns T\ls H\ls I\ls E\ls L\ls E$^1$\break
\and L.\ns J.\ns C\ls U\ls M\ls M\ls I\ls N\ls G\ls S$^2$}

\affiliation{
$^1$Department of Mathematical Sciences, Loughborough University, \\
Leicestershire, LE11 3TU, UK\\[\affilskip]
$^2$Department of Mathematical Sciences and Center for Applied 
Mathematics and Statistics, New Jersey Institute of Technology, 
Newark, NJ 07102, USA}

\pubyear{2012}
\volume{000}
\pagerange{000--000}
\date{?; revised ?; accepted ?. - To be entered by editorial office}

\begin{document}

\maketitle
\begin{abstract}
We study spreading dynamics of nematic liquid crystal droplets within the 
framework of the long-wave approximation. A fourth order nonlinear parabolic partial 
differential equation governing the free surface evolution is derived. The influence 
of elastic distortion energy and of imposed anchoring variations at the substrate 
are explored through linear stability analysis and scaling arguments, which yield 
useful insight and predictions for the behaviour of spreading droplets. This 
behaviour is captured by fully nonlinear time-dependent simulations of three 
dimensional droplets spreading in the presence of anchoring variations that model 
simple defects in the nematic orientation at the substrate.
\end{abstract}

\section{Introduction\mylab{sec:intro}} 

Thin liquid films have surprisingly wide application in our daily life. From industrial 
coating and painting processes to printing, many current technologies require an 
understanding of fluid flows in which one spatial dimension (the film thickness) is 
significantly smaller than the others (typically, the lateral scales over which film 
thickness changes). Under such circumstances one can use systematic 
asymptotic methods based on a small parameter (the representative aspect ratio 
of the film, which characterises the size of the free surface gradients) to simplify 
the full Navier-Stokes governing equations. Expanding the dependent variables 
of interest, such as fluid velocity, pressure, etc., in terms of this small parameter, 
one can obtain a much more tractable system of reduced equations for the 
leading-order quantities. Despite its simplicity, this approach has been used and 
experimentally tested many times, and has been very successful in describing 
the real physics in a wide range of flows.

While plenty of work has been done with Newtonian fluids, this kind of systematic 
asymptotic treatment of flowing thin complex fluids, in particular liquid crystals, is 
still in its infancy (see~\citet{Munch2006}, \citet{Blossey2006} 
and~\citet{Myers2005} for examples of work on non-Newtonian, but not liquid 
crystal, thin film flows). Liquid crystals are anisotropic liquids, which typically 
consist of rod-like molecules. In a nematic phase, the rod-like molecules have no 
positional order, but they self-align to reach long range directional order. Therefore, 
to have a complete description of a nematic liquid crystal (NLC) flow, one needs to 
consider not only the velocity field, but also the orientational director field. In 
experiments on spreading nematic droplets, \citet{Poulard2005} found that NLC 
droplets spreading on a horizontal substrate exhibit a surprisingly rich range of 
instabilities, in the regimes where Newtonian droplets would only spread stably 
(see also~\citet{Delabre2009a} and~\citet{Manyuhina2010}). 

As regards asymptotic (long wavelength) modelling of such flows, 
\citet{BenAmar2001} derived a strongly elastic model to describe the surface 
evolution of strongly-anchored NLCs, work that was extended by 
\citet{Cummings2004} to the weakly-anchored case; while \citet{Carou2007} 
studied the model for blade coating of NLC in two dimensional space in the limit of 
weak elastic effects. An alternative approach based on energetic arguments was 
presented by \citet{Mechkov2009a}. However, these different approaches lead to 
different predictions for the stability of a thin film; a discrepancy that was reconciled 
only very recently~\citep{Lin2013}. We refer the reader to that paper for more 
details, but briefly, \citet{BenAmar2001} and \citet{Cummings2004} employ the 
same stress balance at the free surface of the film as in standard Newtonian flow, 
balancing pressure with capillarity. When this condition is modified to also include 
an elastic stress, results consistent with the energetics-based approach are 
obtained. Moreover, these consistent results indicate that in the case of strong 
anchoring conditions at both, the solid substrate and the free surface of the 
nematic film, such a film is never unstable~\citep{Lin2013}. Note, that in the 
weakly elastic limit of \citet{Carou2007} the effect does not appear at leading 
order and no effect on stability is seen. To summarise: many questions remain to 
be addressed regarding the instability mechanisms in free surface nematic flows. 
No fully consistent ``lubrication'' model for a three dimensional (3D) situation that 
can account for the weak anchoring effects that are crucial for instability has yet 
been proposed or studied.

In this paper, we implement the long wave approximation to derive a model 
describing the three dimensional free surface evolution of a thin film of NLC on a rigid 
substrate. The model incorporates a novel weak anchoring surface energy 
formulation, and shows satisfactory behaviour in the vicinity of a contact line. 
Simple linear stability analysis permits mechanistic insight into how the anchoring 
energy influences the stability of a spreading NLC droplet. 

\section{Model Derivation \mylab{sec:derivation}}

The main dependent variables governing the dynamics of a liquid crystal in the 
nematic phase are the velocity field $\vect{\bar{v}}=(\bar{u},\bar{v},\bar{w})$, and 
the director field $\vect{n}=(n_1,n_2,n_3)$, the unit vector describing the orientation 
of the anisotropic axis in the liquid crystal (an idealised representation of the local 
preferred average direction of the rodlike liquid crystal molecules). The director 
orientation is a function of space and time which, in the limit that director relaxation 
is fast relative to the flow timescale (the limit considered here) is determined by 
minimising a suitably-defined total energy. Molecules like to align locally, a 
preference that is modelled by a bulk elastic (Frank) energy $\bar{W}$, which is 
minimised subject to boundary conditions. In general, a bounding surface is 
associated with a given preferred direction for $\vect{n}$; this preference is known 
as surface anchoring, and is modelled by an appropriate choice of surface energy. 
Anchoring can be tuned by appropriate treatment of a surface and may be either 
weak or strong. The stress tensor for the NLC is a function of the director orientation, 
hence elastic effects can strongly influence the fluid flow, giving rise to behaviour 
that differs markedly from the isotropic Newtonian case.

\subsection{Leslie-Ericksen Equations \mylab{sec:LEequation}}

The flow of nematic liquid crystal may be described by the Leslie-Ericksen 
equations~\citep{Leslie}. Neglecting inertia, and using over-bars to denote 
dimensional variables (dimensionless variables will be without bars), the flow is 
governed by
\bea
\lambda n_i - \pad{\bar{W}}{n_i} + \left(\pad{\bar{W}}{n_{i,j}}\right)_{,j}+\bar{G}_i=0,
\mylab{eq:LE1}\\
-\pad{\bar{\Pi}}{\bar{x}_i} + \bar{G}_k\, \pad{n_{k}}{\bar{x}_i} 
+ \pad{\bar{t}_{ij}}{\bar{x}_j}=0, \mylab{eq:LE2}\\
\frac{\partial \bar{v}_i}{\partial \bar{x}_i} = 0.
\mylab{eq:LE3}
\eea
representing energy, momentum and mass conservation, respectively. Here, 
$\lambda$ is a Lagrange multiplier ensuring that the director ${\vect n}$ is a unit 
vector. The quantities $\bar{W}$, $\bar{G}$ and $\bar{\Pi}$ are defined by
\beq
2\bar{W} = K\left((\bar{\nabla}\cdot\vect{n})^2+|\bar{\nabla}\wedge\vect{n}|^2\right); 
\mylab{eq:W}
\eeq
\beq \bar{G}_i = -\gamma_1\,\bar{N}_i-\gamma_2\,\bar{e}_{ik}n_k, \quad 
\bar{e}_{ij}=\frac{1}{2}\left(\frac{\partial \bar{v}_i}{\partial \bar{x}_j}
+\frac{\partial \bar{v}_j}{\partial \bar{x}_i}\right); 
\mylab{eq:LE4}
\eeq
\beq \bar{N}_i = \dot{n}_i-\bar{\omega}_{ik}\,n_k, \quad 
\bar{\omega}_{ij}=\frac{1}{2}\left(\frac{\partial \bar{v}_i}{\partial \bar{x}_j}
-\frac{\partial \bar{v}_j}{\partial \bar{x}_i}\right); 
\eeq
\beq \bar{\Pi} =\bar{p}+\bar{W}+ \bar{\psi}_g, 
\eeq
where $K$ is an elastic constant (this form of $\bar{W}$ (\ref{eq:W}) exploits 
the widely-used one-constant approximation~\citep{DeGennes1995}), 
$\gamma_1$ and $\gamma_2$ are constant viscosities; an over-dot denotes a 
material (total) time derivative; $\bar{p}$ is the pressure and $\bar{\psi}_g$ is the 
gravitational potential. Finally, $\bar{t}_{ij}$ is the viscous stress tensor, given by 
\beq
\bar{t}_{ij}=\alpha_1n_kn_p\bar{e}_{kp}n_in_j + \alpha_2\bar{N}_i n_j
+\alpha_3 \bar{N}_j n_i
+\alpha_4 \bar{e}_{ij}+\alpha_5 \bar{e}_{ik}n_kn_j +\alpha_6 \bar{e}_{jk}n_kn_i,
\eeq
where $\alpha_i$ are constant viscosities (related to $\gamma_i$ in 
Eq.~(\ref{eq:LE4}) by $\gamma_1=\alpha_3-\alpha_2$, 
$\gamma_2=\alpha_6-\alpha_5$, and to each other by the Onsager relation, 
$\alpha_2+\alpha_3=\alpha_6-\alpha_5$).

\subsection{Nondimensionalization}

We make the usual long-wave scalings to nondimensionalize the governing 
equations
\[
(\bar{x},\bar{y},\bar{z})=(Lx, Ly, \delta Lz), \quad 
(\bar{u}, \bar{v}, \bar{w})=(Uu,Uv,\delta Uw),
\]
\beq
\bar{t}=\frac{L}{U}\,t, \quad 
\bar{p}=\frac{\mu U}{\delta^2 L}\,p, \quad 
\bar{W}=\frac{K}{\delta^2 L^2}\,W,
\eeq
where $L$ is the lengthscale of typical variations in the $x$ and $y$ directions, 
$U$ is the typical flow speed; $\delta=h_0/L\ll 1$ is the aspect ratio of typical 
variations of the film height $h_0$ (a small slope assumption), and 
$\mu=\alpha_4/2$ was chosen as the representative viscosity scaling in the 
pressure, since this corresponds to the usual viscosity in the isotropic case. Our 
choices of $L$ and $U$ are discussed later in the text.

\subsection{Energetics of director field\mylab{sec:energy}}

It has been shown~\citep{BenAmar2001, Cummings2004} that with the above 
scalings, and provided the inverse Ericksen number $K/ (\mu UL) = O(1)$, the 
coupling terms in Eqs.~(\ref{eq:LE1})-(\ref{eq:LE3}) between the energy and 
momentum equations, represented by $\bar{G}$, can be neglected. The energy 
equations then reduce to the appropriate Euler-Lagrange equations for minimising 
the free energy of the film subject to the constraint $\vect{n}\cdot\vect{n}=1$, 
corresponding to the limit of instantaneous relaxation of the director field. 
Imposing the constraint $\vect{n}\cdot\vect{n}=1$ directly we have a director field 
that is a vector on the unit sphere characterised by two angles,
\beq
\vect{n} = (\sin\theta\cos\phi,\sin\theta\sin\phi,\cos\theta),
\eeq
for some functions $\theta(x,y,z,t)$ and $\phi(x,y,z,t)$, which are the usual 
spherical polar angles. 

The leading order bulk elastic energy, under the long-wave scaling, is given by 
\beq
2W = \theta^2_z + \phi^2_z \sin^2\theta + O(\delta).
\mylab{eq:W}
\eeq
The surface energy at the free surface $z=h(x,y,t)$ is denoted by 
$\cG=\cG(\hat{\theta})$ where $\hat{\theta}$ is the conical director orientation at 
the free surface,
\beq
\hat{\theta}=\theta(x,y,h,t).
\mylab{eq:thetahat}
\eeq
The surface energy $\cG$ takes its minimum when the director takes the preferred 
orientation $\hat{\theta}=0$. Within the long-wave approximation this corresponds 
to a director field perpendicular to the free surface: {\it homeotropic} surface 
anchoring. At the substrate $z=0$ we assume strong {\it planar} anchoring, 
$\theta (x,y,0,t)=\pi /2$, with $\phi$ specified. These anchoring assumptions are 
consistent with the experiments of~\citet{Poulard2005} (but not to all experimental
spreading scenarios; in particular our model is not applicable to the case of fully
degenerate planar anchoring at the lower substrate).

We carry out the free energy minimisation directly using a variational principle. The 
total free energy, $J$, consists of bulk and surface contributions. We write
\beq
J = \int^h_0\int_{\Omega}\,\tilde{\cN}\,W\,dS dz + \int_{\Omega}\,\cG\,dS,
\eeq
where $\Omega$ is the domain occupied by the liquid crystal sample in the $x$-$y$ 
plane and $\tilde{\cN} = K/(\mu U L)$ is the inverse Ericksen number. 
We consider the variations induced in $J$ by small variations in the fields 
$\theta$ and $\phi$. The first variations must both vanish at an extremum and the 
sign of the second variations tells us whether or not we have an energy minimum. 
After an integration by parts, the vanishing of the bulk terms in the first variations of 
$J$ leads to
\begin{eqnarray}
\theta_{zz} = \frac{\phi^2_z}{2} \, \sin 2 \theta & \quad & \mbox{in 
$\Omega \cup \{ 0<z<h\}$},
\mylab{eq:ene1}\\
(\phi_z \sin^2\theta)_z = 0 & \quad & \mbox{in $\Omega \cup\{ 0<z<h \}$}.
\end{eqnarray}
At the free surface, the surface energy $\cG$ is independent of the azimuthal angle 
$\phi$ (conical anchoring), hence a natural boundary condition on $\phi$ emerges 
from the surface contribution to the first variation of $J$ with respect to $\phi$: 
$\phi_z\sin^2\theta=0$ on $z=h$. The angle $\phi$ is thus independent of $z$; and 
with our assumption of strong anchoring at the substrate, we then have 
\beq
\phi = \phi (x,y)
\eeq
determined by the imposed substrate anchoring pattern. For $\theta$, 
Eq.~(\ref{eq:ene1}) reduces to $\theta_{zz}=0$, and the strong planar anchoring 
condition is imposed on $z=0$. We then have 
\beq
\theta = a(x,y,t) \, z + \frac{\pi}{2},
\mylab{eq:director}
\eeq
where $a$ is determined by the condition that the surface contribution in the first 
variation vanish, 
\beq
\cG_{\hat{\theta}} + \tilde{\cN}\,a =0.
\mylab{eq:ene2}
\eeq

\subsubsection{Surface Energy \mylab{sec:surfaceenergy}} 

For relatively thick films, the director angle $\theta$ can easily adjust to the 
preferred values at each surface. As the film gets thin, and in particular near 
precursor layers or contact lines, there is a very large energy penalty to pay for 
bending between two fixed angles across a very short distance $h$. In this paper 
we assume the existence of a thin precursor film ahead of a bulk droplet, of 
thickness $0<b\ll 1$ (this is also the case in the experiments 
of~\citet{Poulard2005}). To avoid a near-singularity in the director orientation 
within the precursor, we allow the anchoring to be relaxed as $h\to b$.  

To capture these two limiting behaviours for thick and very thin films, we 
propose that the change in director angle across the fluid layer, $ah$, 
approaches a prescribed value $\Theta$ (the difference in the preferred 
angles at the free surface and solid substrate) as $h\to\infty$; and approaches 
zero as the film thickness $h\to b$. Similar to the approach 
of~\cite{Cummings2011}, we introduce an ad hoc anchoring condition based 
on specifying this change in director angle by 
\beq
ah = \Theta\,m(h),
\mylab{eq:surface1}
\eeq
where $m(h)$ is a monotone increasing function of $h$ with $m(b)=0$ and 
$m(\infty)=1$. With our assumption of homeotropic alignment at the free 
surface (and with the assumed lubrication scalings), $\Theta=-\pi/2$. 

Though we did not specify the surface energy $\cG$ in the above, it is implicitly
imposed, and easily recovered.  Based on the above, the director angle $\theta$ 
is given by $\theta = (\pi/2) (1-z m(h)/h)$, so that the angle $\hat{\theta}$ at the 
free surface, defined by (\ref{eq:thetahat}), is given as a function of $h$ by
\beq
\hat{\theta} = \frac{\pi}{2} (1- m(h) ).
\mylab{eq:surface2}
\eeq
The surface energy must satisfy equation (\ref{eq:ene2}). 
Since Eq.~(\ref{eq:surface2}) is not trivially inverted to give $h(\hat{\theta})$, 
we use the chain rule to obtain
\beq
\frac{d \cG}{d h} =  \cG_{\hat{\theta}} \frac{d\hat{\theta}}{dh} 
=-\cN \,\frac{m(h) m'(h)}{h},
\mylab{eq:surface3}
\eeq
where $\cN = \Theta^2\tilde{\cN}$. 
Equation (\ref{eq:surface3}) defines the surface energy $\cG$ in terms of the 
film height $h$.  The expression in terms of director angle at the free surface, 
$\hat{\theta}$, may be recovered by use of Eq.~(\ref{eq:surface2}).  

Note that Eqs.~(\ref{eq:director}) and (\ref{eq:surface1}) also imply that the bulk elastic 
energy in Eq.~(\ref{eq:W}) becomes
\beq
W = \frac{\Theta^2}{2}\frac{m^2}{h^2}.
\eeq
As a result we have the contribution of nematic elasticity to the free energy as
\beq
J = \int_{\Omega}\,\left[\frac{\cN}{2}\,\frac{m^2}{h}  + \cG\, \right]dS.
\mylab{eq:JJ}
\eeq

\subsection{Momentum Equation}

For the momentum equations, Eqs.~(\ref{eq:LE2}), balancing dominant terms gives 
\[
\pad{\Pi}{x}\sim \pad{t_{13}}{z}, \quad \pad{\Pi}{y}\sim \pad{t_{23}}{z},
\]
in dimensionless form. Based on the long-wave scalings, to leading order we have 
\beq
t_{13} = (A_1+A_2\cos 2\phi)\, u_z + A_2\sin 2\phi\, v_z,\quad
t_{23} = A_2\sin 2\phi\,u_z + (A_1-A_2\cos 2\phi) \,v_z,
\eeq
where 
$A_1= 1+ (\alpha_5-\alpha_2)\,\cos^2\theta 
+ \alpha_1\sin^2\theta\cos^2\theta + (\alpha_3+\alpha_6)\sin^2\theta/2$, 
$A_2= \alpha_1\sin^2\theta\cos^2\theta + (\alpha_3+\alpha_6)\sin^2\theta/2$, 
and the 
$\alpha_i$ are normalised by $\mu=\alpha_4/2$. As a result, the leading order 
equations are 
\begin{eqnarray}
\pad{p}{x} + \tilde{\cN}\theta_z\theta_{zx} &=& \frac{\partial}{\partial z}
\left\{(A_1+A_2\cos 2\phi)\, u_z + A_2\sin 2\phi\, v_z\right\},\mylab{eq:LCm1}\\
\pad{p}{y} + \tilde{\cN}\theta_z\theta_{zy} &=& \frac{\partial}{\partial z}
\left\{A_2\sin 2\phi\,u_z + (A_1-A_2\cos 2\phi) \,v_z\right\},\mylab{eq:LCm2}\\
\pad{p}{z} &=& -\cB,\mylab{eq:LCm3}
\end{eqnarray}
where $\cB=\delta^3 \rho g L^2/{\mu U}$ is the Bond number.

We assume that the normal component of the stress at the free surface balances 
surface tension (the isotropic component of the surface energy, $\gamma$) times 
curvature, and that the in-plane component of the stress is balanced by surface 
tension (surface energy) gradients in the plane of the surface. This yields the 
leading order boundary conditions:
\begin{eqnarray}
p+ \tilde{\cN}\,\theta^2_z &=& -\cC\nabla^2 h, 
\mylab{eq:LCmb1}\\
- \tilde{\cN}(\theta_x\theta_z+\theta^2_z h_x)+
(A_1+A_2\cos 2\phi)\, u_z + A_2\sin 2\phi\, v_z &=& \tilde{\cN} \cG_x,
\mylab{eq:LCmb2}\\
- \tilde{\cN}(\theta_y\theta_z+\theta^2_z h_y)+
A_2\sin 2\phi\,u_z + (A_1-A_2\cos 2\phi) \,v_z &=& \tilde{\cN} \cG_y, 
\mylab{eq:LCmb3}
\end{eqnarray}
where $\cC=\delta^3 \gamma/\mu U$ is an inverse capillary number. Furthermore, 
using Eqs.~(\ref{eq:ene2})-(\ref{eq:surface2}), the equations of the tangential 
stress balances, Eqs.~(\ref{eq:LCmb2})-(\ref{eq:LCmb3}), reduce to $u_z=0$ and 
$v_z=0$. 

We solve Eqs.~(\ref{eq:LCm3})-(\ref{eq:LCmb1}) for $p$ 
\beq
p=\cB(h-z) - \tilde{\cN} a^2 - \cC \nabla^2 h,
\eeq
and substitute in Eqs.~(\ref{eq:LCm1})-(\ref{eq:LCm2}) to obtain $u_z$ and $v_z$ 
using the boundary conditions derived above:
\bes
D\,u_z &=& \left[(A_1 - A_2 \cos 2\phi ) \, (p_x + \tilde{\cN} aa_x)
       - A_2\sin 2\phi \, (p_y + \tilde{\cN} aa_y)\right](z-h), \\
D\,v_z &=& \left[(A_1+A_2 \cos 2\phi) \, (p_y + \tilde{\cN} aa_y)
       - A_2\sin 2\phi \, (p_x + \tilde{\cN} aa_x)\right](z-h),
\ees
where $D=A^2_1-A^2_2$. Finally, using conservation of mass together with the 
relations
\[
\int^h_0 u\,dz = \int^h_0 u_z(h-z)\,dz, \quad \int^h_0 v\,dz = \int^h_0 v_z(h-z)\,dz,
\]
we obtain a partial differential equation governing the evolution of the film height:
\beq
h_t + \nabla\cdot\left[\left\{
\mathfrak{f}_1\,I
+
\mathfrak{f}_2\,
\left[\begin{array}{cc}
\cos 2\phi & \sin 2\phi \\
\sin 2\phi & -\cos 2\phi
\end{array}\right]
\right\}
\cdot\nabla
\left(\cC\nabla^2h-\cB h + \frac{\tilde{\cN}}{2} a^2\right)
\right]=0, 
\mylab{eq:NLCT1}
\eeq
where $I$ is the identity matrix and 
\beq
\mathfrak{f}_1 = \int^h_0\,\frac{A_1}{A^2_1-A^2_2}\,(h-z)^2\,dz, \quad 
\mathfrak{f}_2 = \int^h_0\,\frac{-A_2}{A^2_1-A^2_2}\,(h-z)^2\,dz.
\mylab{eq:NLCT2}
\eeq

Equations~(\ref{eq:NLCT1})-(\ref{eq:NLCT2}) represent a formidable analytical 
challenge. We simplify by approximating the integral expressions using the 
two-point trapezoidal rule, as 
\beq
\mathfrak{f}_1 = \lambda\,h^3,\quad 
\mathfrak{f}_2 = \nu\,h^3,\quad
\lambda = \frac{2+\alpha_3+\alpha_6}{4(1+\alpha_3+\alpha_6)}, \quad
\nu = -\frac{\alpha_3+\alpha_6}{4(1+\alpha_3+\alpha_6)}.
\eeq
For $-1<\alpha_3+\alpha_6<0$ (which is the case for all common nematic liquid 
crystals), we have $\lambda>\nu>0$. By including these quantities and our 
chosen surface energy $\cG$ from \S~\ref{sec:surfaceenergy}, the equation can 
be rewritten as
\beq
h_t + \nabla\cdot\left[
h^3\tilde{\nabla}\left(\cC\nabla^2h-\cB h\right)
+\cN\left(mm'h-m^2\right)\tilde{\nabla}h
\right]=0, 
\mylab{eq:NLC1}
\eeq
where $\cN=\Theta^2 K/ \mu U L$ and 
\beq
\tilde{\nabla} = \left(\lambda I+\nu\,\left[
\begin{array}{cc}
\cos 2\phi & \sin 2\phi \\
\sin 2\phi & -\cos 2\phi
\end{array}
\right]\right)\cdot\nabla.
\mylab{eq:NLC2}
\eeq

\subsection{Model Summary}

Our final model consists of the partial differential equation, Eq.~(\ref{eq:NLC1}), 
where $\tilde{\nabla}$ is defined in Eq.~(\ref{eq:NLC2}) with the anchoring 
condition at the free surface, $m(h)$, and the anchoring pattern at the substrate, 
$\phi(x,y)$, to be specified. We have five dimensionless positive parameters: 
$\lambda$, $\nu$, $\cC$, $\cB$, $\cN$, giving a solution space that is potentially 
very large. In the following analysis and simulations, we assume a balance 
between surface tension and gravity, setting $\cC=\cB=1$, meaning physically 
that the typical length scale $L$ considered is the capillary length, 
$\sqrt{\gamma/\rho g}$. Clearly, alternative choices can be made if a different 
balance, or different lengthscales, are to be considered.

On the other hand, it has been shown~\citep{Lin2013} that the evolution equation 
for a NLC film in the limit of strong anchoring can be written in a variational or 
gradient dynamics form in line with such formulations for films of simple liquids 
(see, e.g., \citet{S.M.1993,Thiele2010}), films of mixtures~\citep{Thiele2011} and 
surfactant-covered films~\citep{Thiele2012}. We would like to point out that this is also 
true for the current model. In such a formulation, the evolution of the film thickness 
$h$ follows a dissipative gradient dynamics governed by the equation
\beq
h_t = \nabla\cdot\left[Q(h)\tilde{\nabla}\left(\frac{\delta F}{\delta h}\right)\right],
\mylab{eq:GD1}
\eeq
where $Q(h)$ is the mobility function and $F$ is the free energy functional written 
as
\beq
F[h] = \int_{\Omega}\,\left[\cC\left(1 + \frac{(\nabla h)^2}{2}\right) 
+ \frac{\cB}{2}\,h^2\right]\,dS+J.
\mylab{eq:GD2}
\eeq
(The contribution of nematic elasticity on the free energy functional, $J$, 
is given in Eq.~(\ref{eq:JJ}).) 
By introducing $F$ into Eq.~(\ref{eq:GD1}) and noting that the surface energy 
$\cG$ is coupled with the film thickness through Eq.~(\ref{eq:surface3}), we 
obtain the evolution equation
\beq
h_t = \nabla\cdot\left[Q(h)\tilde{\nabla}\left(
-\cC \nabla^2h + \cB h - \frac{\cN}{2}\frac{m^2}{h^2}
\right)\right].
\mylab{eq:GD3}
\eeq
One should note that Eq.~(\ref{eq:NLC1}) and Eq.~(\ref{eq:GD3}) are identical 
when $Q(h)=h^3$.

\section{Analysis and Results \mylab{sec:aar}}

In this Section we investigate some limiting cases of the model analytically, and 
carry out additional simulations for spreading nematic films and droplets in selected 
flow configurations. The time-dependent simulations that we report below are based on 
an Alternative-Direction-Implicit (ADI) method (as outlined by~\citet{Witelski2003}) 
with variable time stepping based on a Crank-Nicolson scheme; 
see~\citet{Lin2012a} for further details.

\subsection{Influence of anchoring patterns at the substrate}

To gain some insight into our model we first compare two different uni-directional 
substrate anchoring patterns, in the simple case where flow is independent of $y$, 
and the fluid spreads uniformly in the $x$ direction. Assuming $\phi=0$ (director 
orientation at the substrate parallel to the fluid flow direction), Eq.~(\ref{eq:NLC1}) 
becomes
\beq
h_t + (\lambda+\nu)\, \partial_x\left[h^3\,(h_{xxx}-h_x)+\cN\left(mm'h-m^2\right)
h_x \right]=0.
\mylab{eq:par1}
\eeq
If, on the other hand, we assume $\phi=\pi/2$, so that the substrate director 
orientation is perpendicular to the fluid flow (but still in the plane of the substrate), 
Eq.~(\ref{eq:NLC1}) becomes
\beq
h_t + (\lambda-\nu)\, \partial_x\left[h^3\,(h_{xxx}-h_x)+
\cN\left(mm'h-m^2\right)h_x \right]=0.
\mylab{eq:par2}
\eeq
Equations~(\ref{eq:par1}) and (\ref{eq:par2}) are identical once time is rescaled by 
a constant. In this simple example, the substrate pattern only affects the flow 
timescale, effectively making the fluid viscosity
$\phi$-dependent. The NLC flows faster 
when the anchoring pattern is parallel to the flow direction (effective viscosity is 
smaller), and slower when it is perpendicular (effective viscosity is larger).

To focus more on the influence of the anchoring patterns at the substrate, we next
consider a weak conical free surface anchoring on $\theta$ so that the director 
orientation is mainly determined by the strong planar anchoring at the substrate, 
i.e., $\theta\equiv \pi/2$ and $m(h) \equiv 0$. The contribution of nematic bending 
elasticity then disappears. In particular, the integral expressions in 
Eq.~(\ref{eq:NLCT2}) can now be evaluated exactly; there is no need to 
approximate them as was done to obtain Eq.~(\ref{eq:NLC1}). The resulting 
equation is 
\beq
h_t + \nabla\cdot\left[\frac{2h^3}{3}
\tilde{\nabla}\,
\left(\nabla^2h- h  \right) \right]=0.
\mylab{eq:NLCW}
\eeq

Figure~\ref{fig:stripe} shows the solution of Eq.~(\ref{eq:NLCW}) computed using 
ADI-based simulations of a configuration where the anchoring imposed at the 
substrate appears as a striped pattern, as shown in Fig.~\ref{fig:stripe}(a): 
$\phi=\pi/2$ for $x\in (4n-1,4n+1)$, $n=0$, $\pm 1$ and $\pm 2$, and $\phi=0$ 
otherwise. The initial film profile, shown as a surface contour plot in 
Fig.~\ref{fig:stripe}(b), is taken as
\beq
h(x,y,0)=0.45\tanh(-5(y-10))+0.55, 
\eeq
with the front position being a straight line parallel to the $x$-axis, and spreading 
in the $+y$ direction. The computational domain is defined by $-L_x \le x \le L_x$ 
and $0 \le y \le L_y$, with $L_x = 10$ and $L_y = 20$. The implemented boundary 
conditions are 
$h_x(\pm L_x, y, t)=h_y(x, 0, t)=h_y(x, L_y, t)=h_{xxx}(\pm L_x, y, t)=h_{yyy}(x, 0, t)=0$,  
$h(x, L_y, t)=b$, where $b$ is the thickness of the prewetting layer (precursor film) 
used to remove the contact line singularity.  In the simulation scenarios that follow 
(in both this and the following Section), the exact value given to $b$ was found to 
have only a weak influence on spreading (with faster spreading for larger $b$), but 
no influence of the exact value of $b$ on the stability of the flow has been found. 
We use $b=0.1$ here, and on uniform computational grids specified by the grid 
spacing $\delta x = \delta y  = 0.1$; these values are found to be sufficient to 
guarantee numerical convergence.

\begin{figure}
\centering
\subfigure[front position]{\includegraphics[width=1.5in]{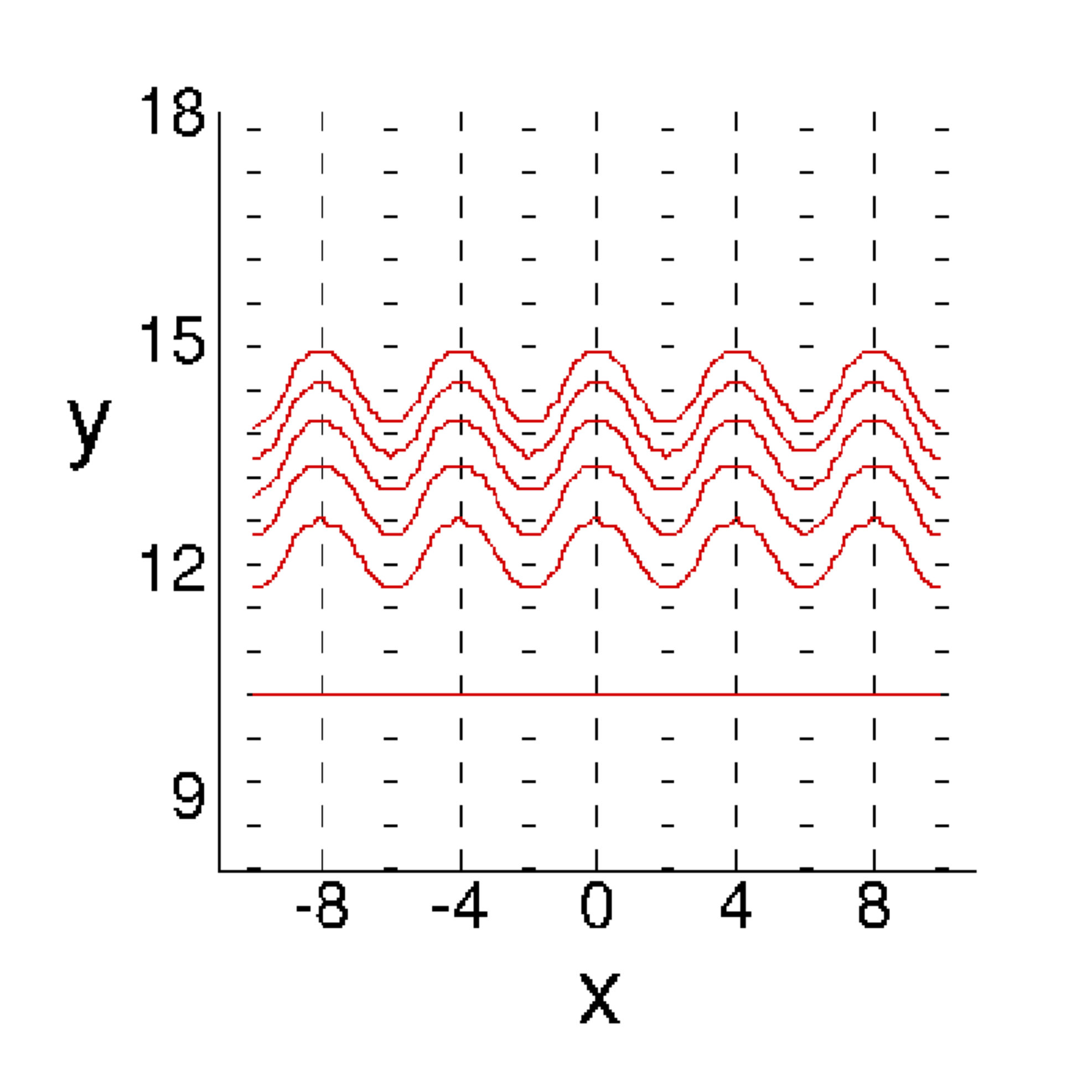}}
\subfigure[$t=0$]{\includegraphics[width=1.2in]{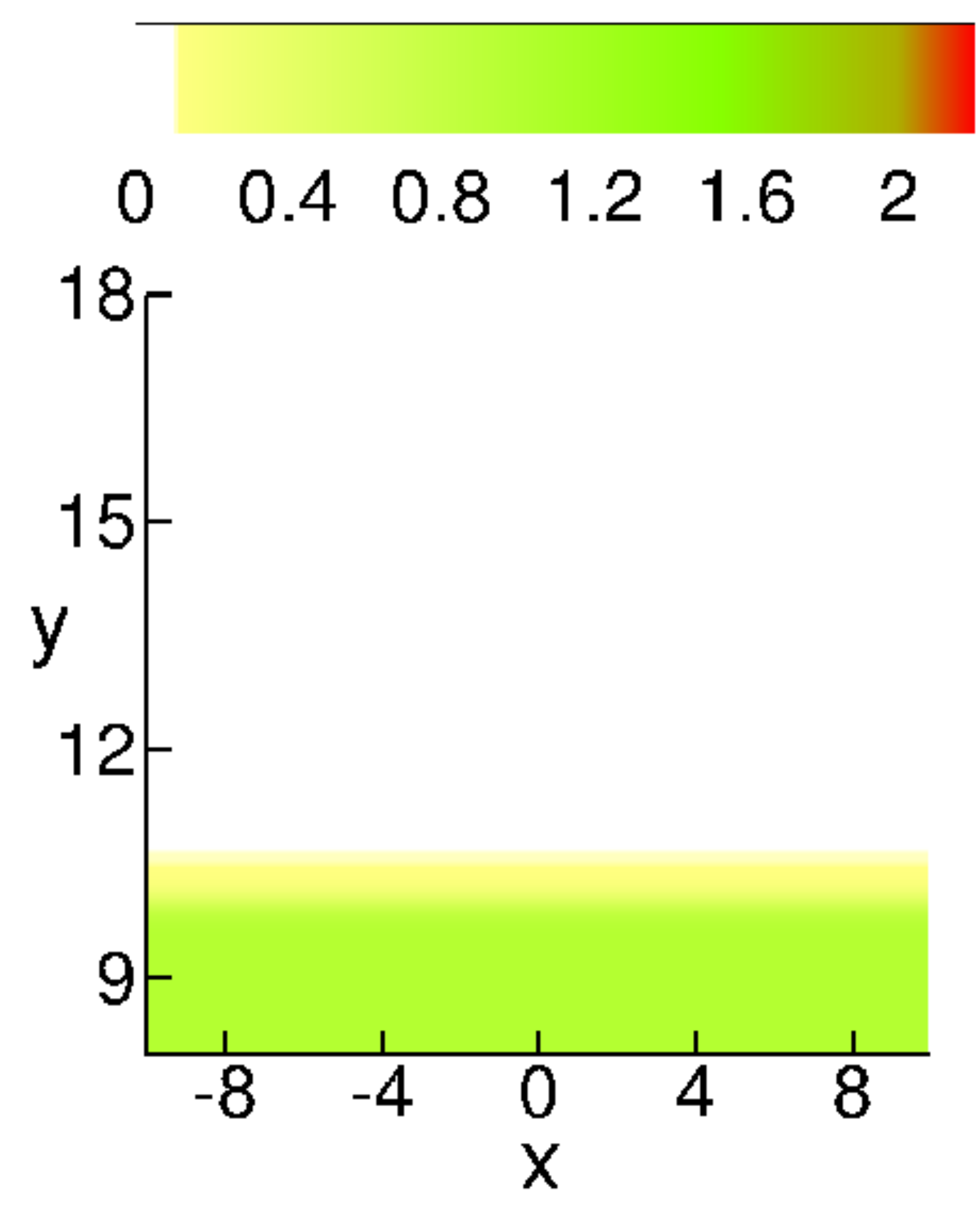}}
\subfigure[$t=50$]{\includegraphics[width=1.2in]{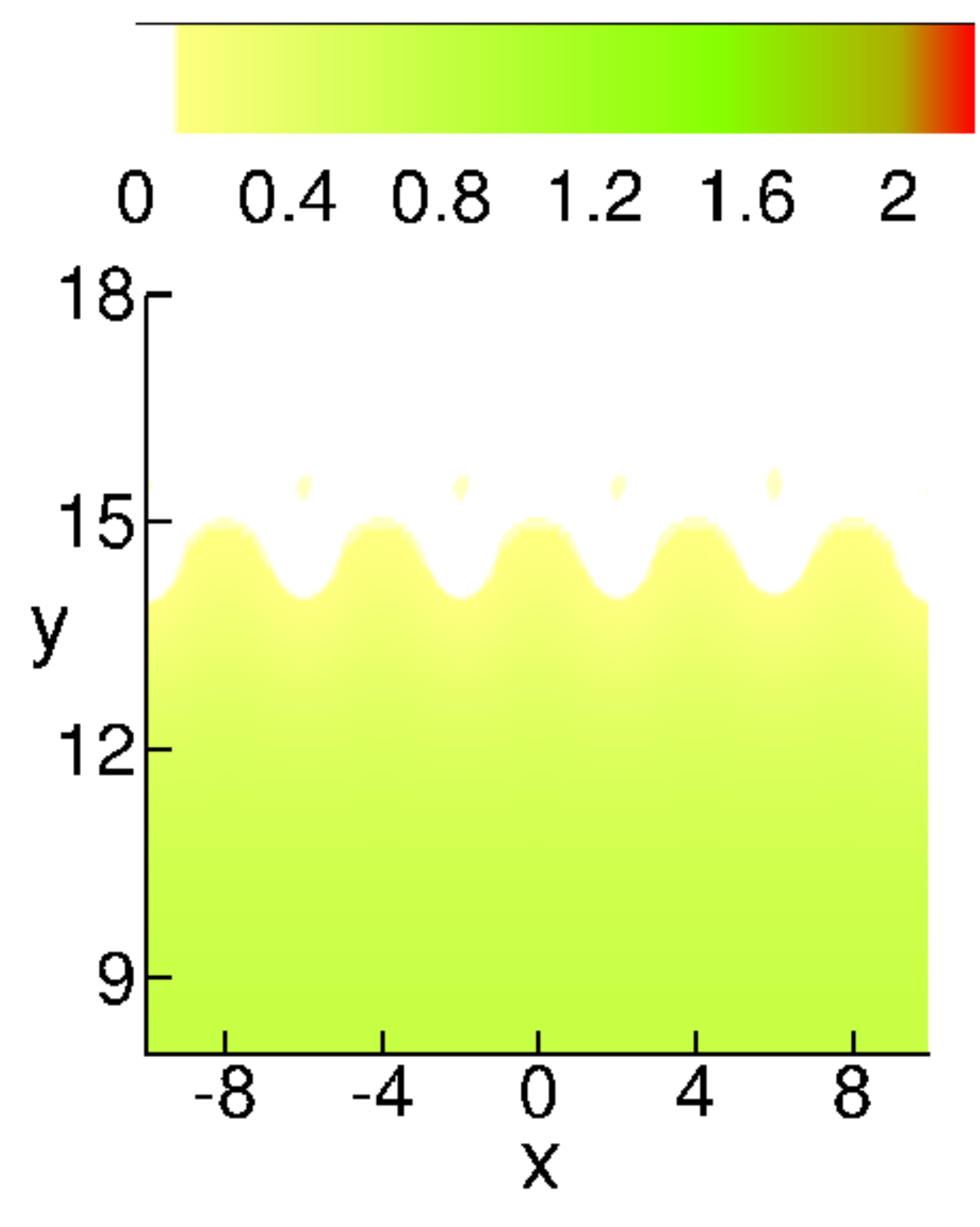}}
\subfigure[$y$ cross section]{\includegraphics[width=1.3in]{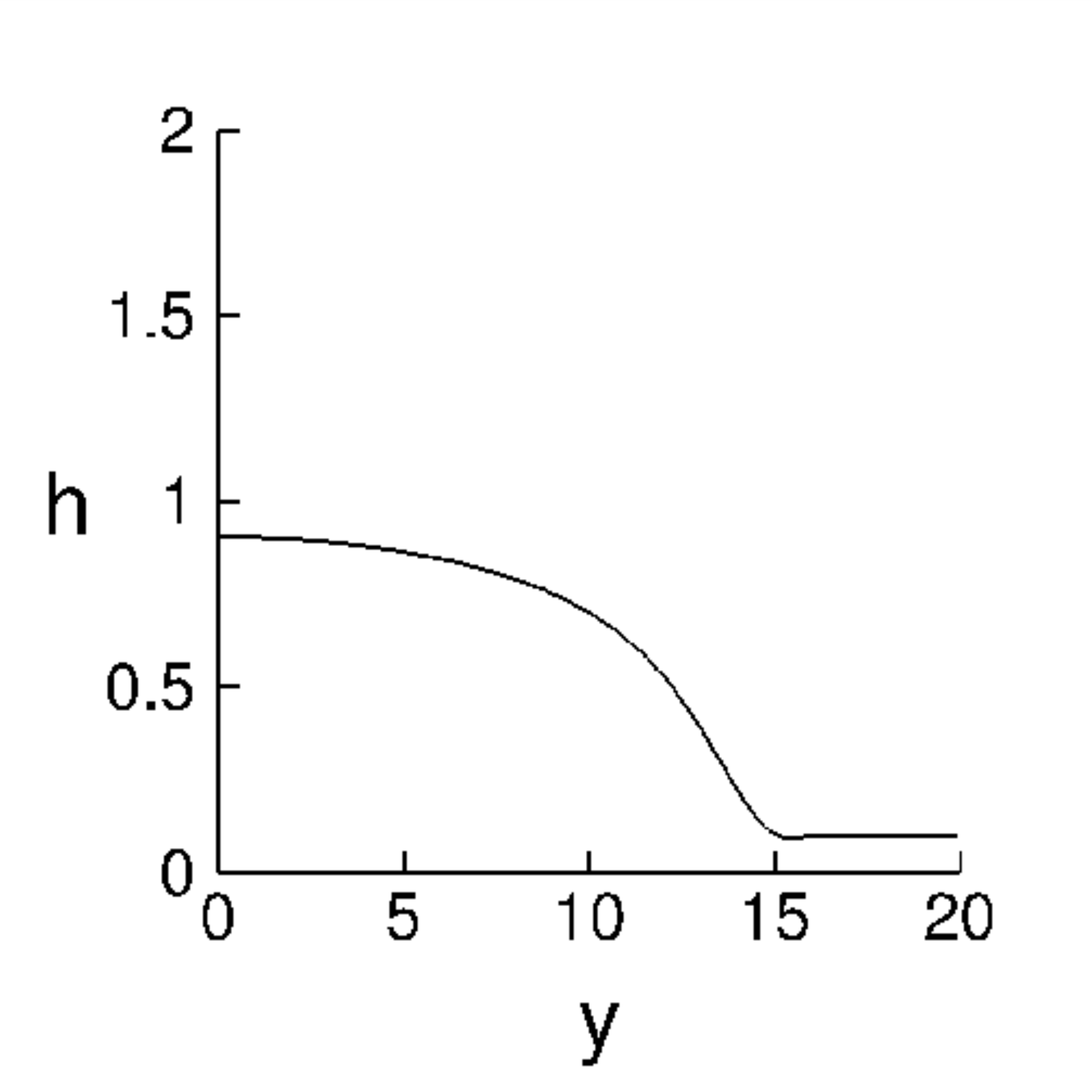}}
\caption{(Colour online) Spreading NLC film on a stripe-patterned substrate 
($\lambda=1$, $\nu=0.5$, $\cN=0$). 
(a) The dashed (black) lines indicate the anchoring at the substrate. The solid (red) 
curves show the front position at $\Delta t=10$ time intervals between successive 
curves. The initial front position is shown as a straight (red) line at $y\approx 10$. 
(b - c) Surface contour plot of the film at $t=0$, $50$, respectively. (d) Cross 
section of the film, $h(x=0, y, t=50)$. 
}
\mylab{fig:stripe}
\end{figure}

Figure~\ref{fig:stripe}(a) shows the evolution of the spreading fluid front. The 
anchoring pattern at the substrate clearly influences the spreading: as 
Fig.~\ref{fig:stripe}(a) shows, in line with the observations discussed above, the 
front moves fastest when the fluid motion is parallel to the anchoring pattern, and 
slowest when flow is perpendicular to anchoring. As the anchoring changes 
periodically, the speed of the front  transitions between the two extremes, giving 
rise to a sawtooth pattern. It should be noted that although Eqs.~(\ref{eq:par1}) 
and (\ref{eq:par2}) are indicative of different wavespeeds for an isolated moving 
front, the amplitude of the front perturbation does not increase linearly in time, as 
seen in Fig.~\ref{fig:stripe}(a). Instead, the amplitude approaches a constant value. 
This constant amplitude is determined by the balance between different effective 
viscosities and surface tension effects.

Figure~\ref{fig:stripe}(c) shows a surface contour plot of the profile at a late time, 
$t=50$, and Figure~\ref{fig:stripe}(d) shows its cross section at $x=0$. Note the 
absence of a capillary ridge behind the front, indicating the stability of the 
underlying flow. The sawtooth pattern that develops in the spreading front here, 
though reminiscent of a fingering instability, is no such thing: it is simply the result 
of the anchoring inhomogeneity imposed at the substrate.

\subsection{Influence of anchoring condition at the free surface}

We now analyse the effect of free surface anchoring on the director angle, 
$\theta$. We begin by reviewing the linear stability analysis (LSA) of a simple flat 
film of height $h=h_0$ in the 2D case in which variations with respect to the $y$ 
coordinate are neglected, so that both director field and flow are confined to the 
$(x,z)$-plane~\citep{Cummings2011}. This analysis is found in practice to give a 
remarkably good indication regarding stability of spreading 3D droplets, 
considered in \S\ref{sec:numerics} below. In that section we present several 
simulations of stable/unstable 3D droplet evolution of a chosen surface anchoring 
function to illustrate the kind of behaviour that our model can reproduce in the 
presence of some simple substrate anchoring patterns.

\subsubsection{Linear Stability Analysis of a flat film \mylab{sec:LSA}}

With the director confined to the $(x,z)$ plane, $\phi \equiv 0$, and no 
$y$-dependence, Eq.~(\ref{eq:NLC1}) reduces to Eq.~(\ref{eq:par1}). Assuming 
$h=h_0+\xi$ and $|\xi|\ll h_0$ in this equation, we find
\beq
\xi_t + (\lambda+\nu) h_0^3\left[
\xi_{xxxx}-\xi_{xx}+\cN M(h_0) \xi_{xx}
\right]=0, 
\eeq
where
\beq
M(h) = \frac{m(h) m'(h) h -m(h)^2}{h^3}.
\mylab{eq:capm}
\eeq
By setting $\xi\propto \exp{(ikx+\omega t)}$, we obtain the dispersion relation
\beq
\omega = -(\lambda+\nu) h^3_0\left[k^4+(1-\cN M(h_0))k^2\right].
\mylab{eq:dispersion}
\eeq
The flat film is thus unstable to sufficiently long-wavelength perturbations if 
$\cN M(h_0)>1$. When this is the case, perturbations with wavenumbers 
$k\in (0,k_c)$ are unstable, where $k_c=\sqrt{\cN M(h_0)-1}$ is the critical 
wavenumber. The fastest-growing wavenumber (for which the growth rate is 
the largest) is $k_m= {k_c/\sqrt{2}}$, corresponding to the wavelength
\beq
l_m = \frac{2\pi}{k_m} = \frac{2\pi}{\sqrt{(\cN M(h_0)-1)/2}},
\eeq
and the growth rate $\omega_m=(\lambda+\nu) h^3_0(\cN M(h_0)-1)^2/4$.

\subsubsection{Strong surface anchoring }

We consider firstly a strong homeotropic anchoring, given by $m\equiv 1$. In this 
limit, the evolution equation becomes
\beq
h_t + \nabla\cdot\left[
h^3\tilde{\nabla}\left(\cC\nabla^2h-\cB h\right)
-\cN\tilde{\nabla}h
\right]=0, 
\mylab{eq:NLC3}
\eeq
Note that, the elastic contributions to the governing equation are purely diffusive. 
A version of this limit was derived via alternative energetic considerations 
by~\citet{Mechkov2009a} (see also \citet{Lin2013} for a more in-depth discussion
of the strong anchoring case). Although the director field 
corresponding to strong anchoring becomes singular as the film height $h\to 0$, 
the PDE governing the film height in this limit is well-behaved and will never 
exhibit an instability. This observation suggests that the weak free surface anchoring, 
necessary on physical grounds for the director to be nonsingular as the film height 
goes to zero, is key for the instability mechanism.

\subsubsection{Weak surface anchoring }

For a weak surface anchoring, there are many possible forms for $m(h)$ that 
satisfy our basic requirement $m(b)=0$, $m(\infty)=1$. Here, as an example, 
we take
\begin{equation}
m(h)= f (h;b)
\left(\frac{h^{\alpha}}{h^{\alpha}+\beta^{\alpha}}\right).
\mylab{eq:numanc}
\end{equation}
where $\alpha$ and $\beta$ are positive constants that tune the relaxation of the
anchoring for film heights larger than the precursor, and $f(h;b)$ provides the ``cutoff'' 
behavior as the precursor is approached.  In the simulations presented in this paper
we choose $f(h;b)=[ \tanh((h-2b)/w)+1]/2$, where $w$ fixes the size of the $h$-range
over which $m(h)$ is turned off as $h\to b$ ($w\to 0$ gives a simple discontinuous
switch; we assign a small positive value, $w=0.05$, to smooth this behaviour).
This choice for $m(h)$ ensures that the director field for thin films, roughly less than 
$b$, lies in the plane $\theta =\pi/2$, with $\phi$ dictated by the substrate anchoring 
conditions. We note that the exact 
functional form given to $m(h)$ does not influence the results to any significant
degree, as long as $m(h)$ changes sufficiently rapidly for $h\sim b$.

Figure~\ref{fig:mh} (a) shows the anchoring condition at the free surface, 
Eq.~(\ref{eq:numanc}), with $\alpha=\beta=1$ and $b=0.1$. It can be seen that 
the anchoring condition approaches $0$ for thin films and increasingly 
approaches $1$ when the film thickness gets thicker. Figure~\ref{fig:mh} (b) 
shows the function $\cN m(h)-1$ for $\cN=0.2$ (dashed (black) curve) and for 
$\cN=1$ (solid (blue) curve). 
As demonstrated in \S\ref{sec:LSA}, Eq.(\ref{eq:dispersion}), the flat film with 
thickness $h_0$ is unstable if 
$\cN m(h_0)-1>0$. One can see that for $\cN=0.2$, flat films are always stable 
while for $\cN=1$, there exists a range of film thicknesses (the critical values are 
marked by circles (red) in Fig.~\ref{fig:mh} (b)) that exhibit instabilities. 

\begin{figure}
\centering
\subfigure[$m(h)$]{\includegraphics[height=1.7in]{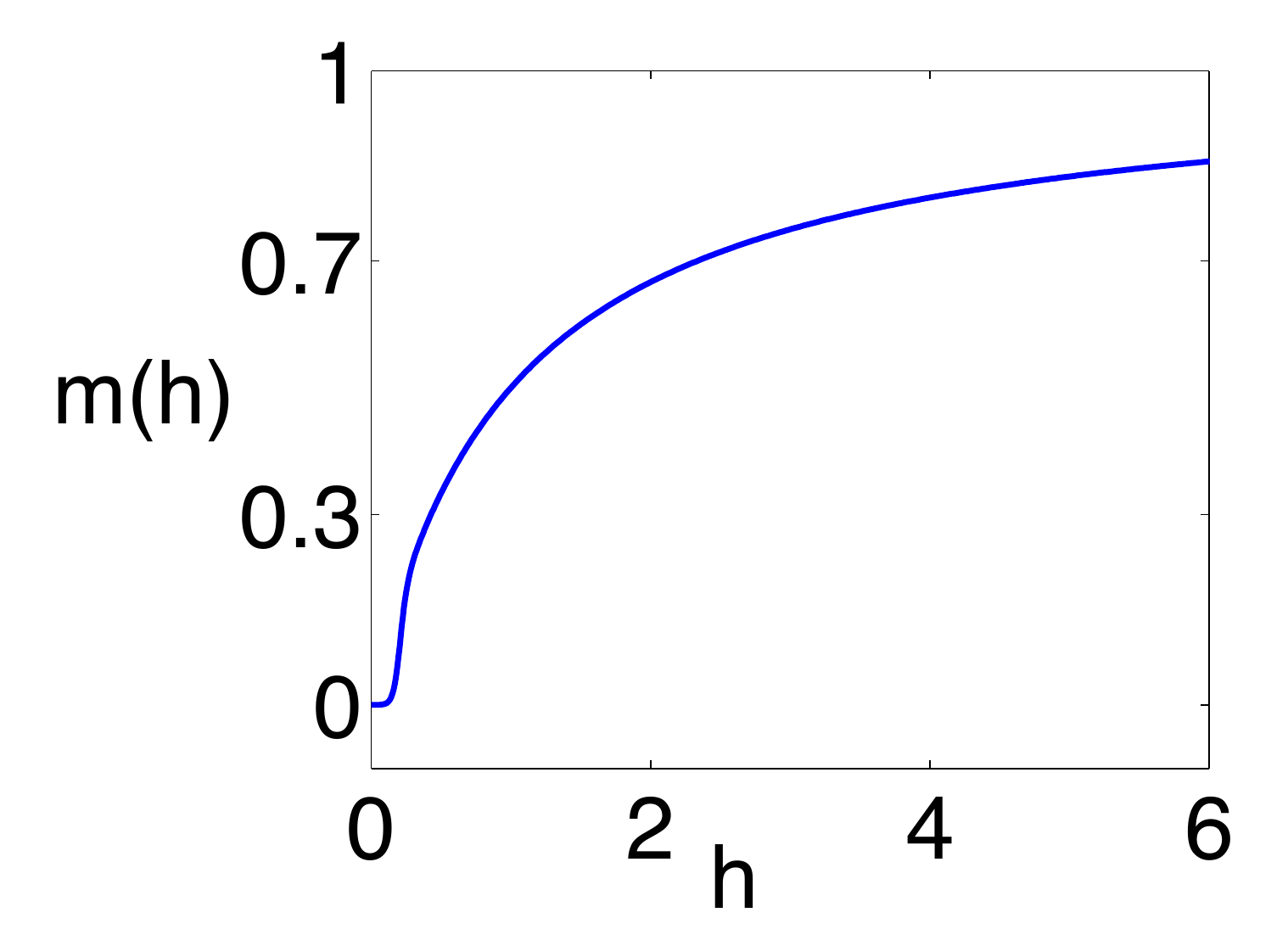}}
\subfigure[$NM(h)-1$]{\includegraphics[height=1.7in]{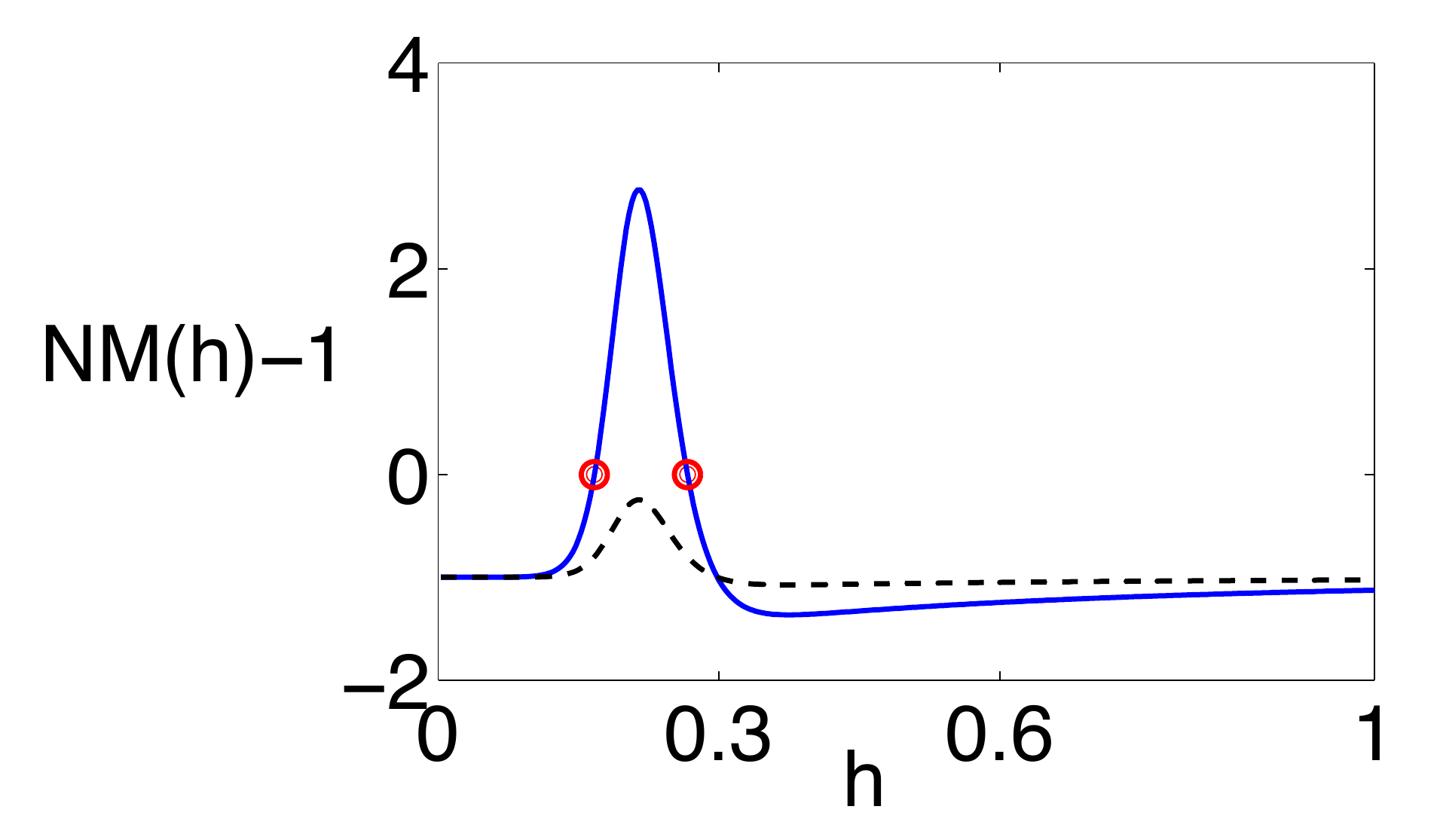}}
\caption{(Colour online) 
(a) The anchoring condition, $m(h)$, as defined in Eq.~(\ref{eq:numanc}) with 
$\alpha=\beta=1$ and $b=0.1,\,w=0.05$. (b) $\cN M(h)-1$ as a function of $h$ where 
$M(h)$ is defined in Eq.~(\ref{eq:capm}). The solid (blue) curve is for $\cN=1$ 
and the dashed (black) curve for $\cN=0.2$. The red circles indicate the critical 
values where $\cN M(h)-1=0$.
}
\mylab{fig:mh}
\end{figure}

\subsection{Numerical results \mylab{sec:numerics}}

In this section we present several simulations of three dimensional spreading 
nematic droplets, in which the influence of the anchoring condition at the substrate 
can be directly investigated. In particular, as test cases we consider spreading on 
substrate anchoring patterns that mimic the director structure near different types 
of nematic defects, in order to analyse the influence of local director structure on 
spreading (see~\citet{LIN2012} for a related analysis of strictly 2D ``defects''). 
Such defects are classified according to their topological winding number $s$: as 
a small planar circuit around the defect is traversed exactly once, the director field 
rotates through an angle $2\pi s$. Specifically, we have 
\begin{equation}
\phi(x, y) = s\,\tan^{-1}\left(\frac{y}{x}\right).
\mylab{eq:numphi}
\end{equation}
While this description cannot capture the true physics very close to the defect
centre (the director field notion breaks down there and one has to introduce a
tensorial order parameter for a detailed description; see e.g.~\citet{DeGennes1995} 
for a discussion) we suggest that it may provide a reasonable 
description of the macroscopic free surface evolution in the presence of a pinned 
defect. More importantly, for different choices of $s$, Eq.(\ref{eq:numphi}) provides 
examples of generic surface anchoring patterns that provide useful demonstrations 
of our model behaviour.

We solve numerically Eq.~(\ref{eq:NLC1}) on such an anchoring pattern, our choice 
of parameters guided by the LSA results presented in \S~\ref{sec:LSA} above. The 
computational domain in all cases is chosen as $-L_x \le x \le L_x$ and 
$-L_y \le y \le L_y$ with $L_x = L_y = 20$, and with the boundary conditions
\beq
h(x,\pm L_y,t)=h(\pm L_x, y,t)=b, \quad h_y(x,\pm L_y,t)=h_x(\pm L_x, y,t)=0.
\eeq
For our first set of simulations we take as initial condition a smoothed cylinder of 
radius $10$ and height $h_0=0.2$, with a precursor film of thickness $b=0.05$ 
covering the rest of the domain. While the exact functional form for $h(x,y,0)$ only 
weakly influences the consequent evolution, we give it here for definiteness:
\beq
h(x,y,0)=\frac{h_0-b}{2}\tanh\left(-(\sqrt{x^2+y^2}-10)\right)+\frac{h_0+b}{2}.
\mylab{eq:init}
\eeq
Finally, the parameters appearing in Eq.~(\ref{eq:NLC2}), (\ref{eq:numanc}) are 
chosen as $\alpha=1$, $\beta=1$, $\lambda=1$ and $\nu=1/2$. The value of 
$\cN$ is given in each figure caption.

Figure~\ref{fig:n02qnh}(b) and (c) shows the evolution of a stably spreading 
nematic droplet, for $\cN=0.2$. The anchoring at the substrate mimics the director 
structure near a defect of type $s=-1/2$, shown in Figure~\ref{fig:n02qnh}(a). 
Due to the non-uniformity of this pattern in the radial direction, the droplet spreads 
asymmetrically, but in the manner that might be expected for the prescribed 
anchoring pattern.  Consistent with the results of our LSA in \S~\ref{sec:LSA} we do 
not observe any free surface instabilities: the corresponding flat film is
stable for the chosen parameters.

\begin{figure}
\centering
\subfigure[anchoring condition]{\includegraphics[width=2.0in]{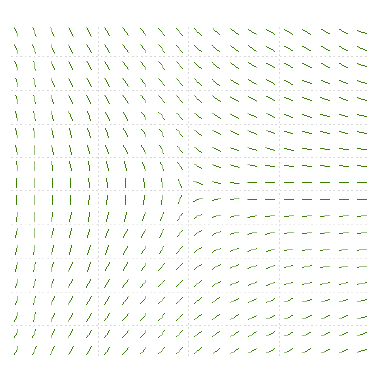}}
\subfigure[$t=0$]{\includegraphics[width=1.5in]{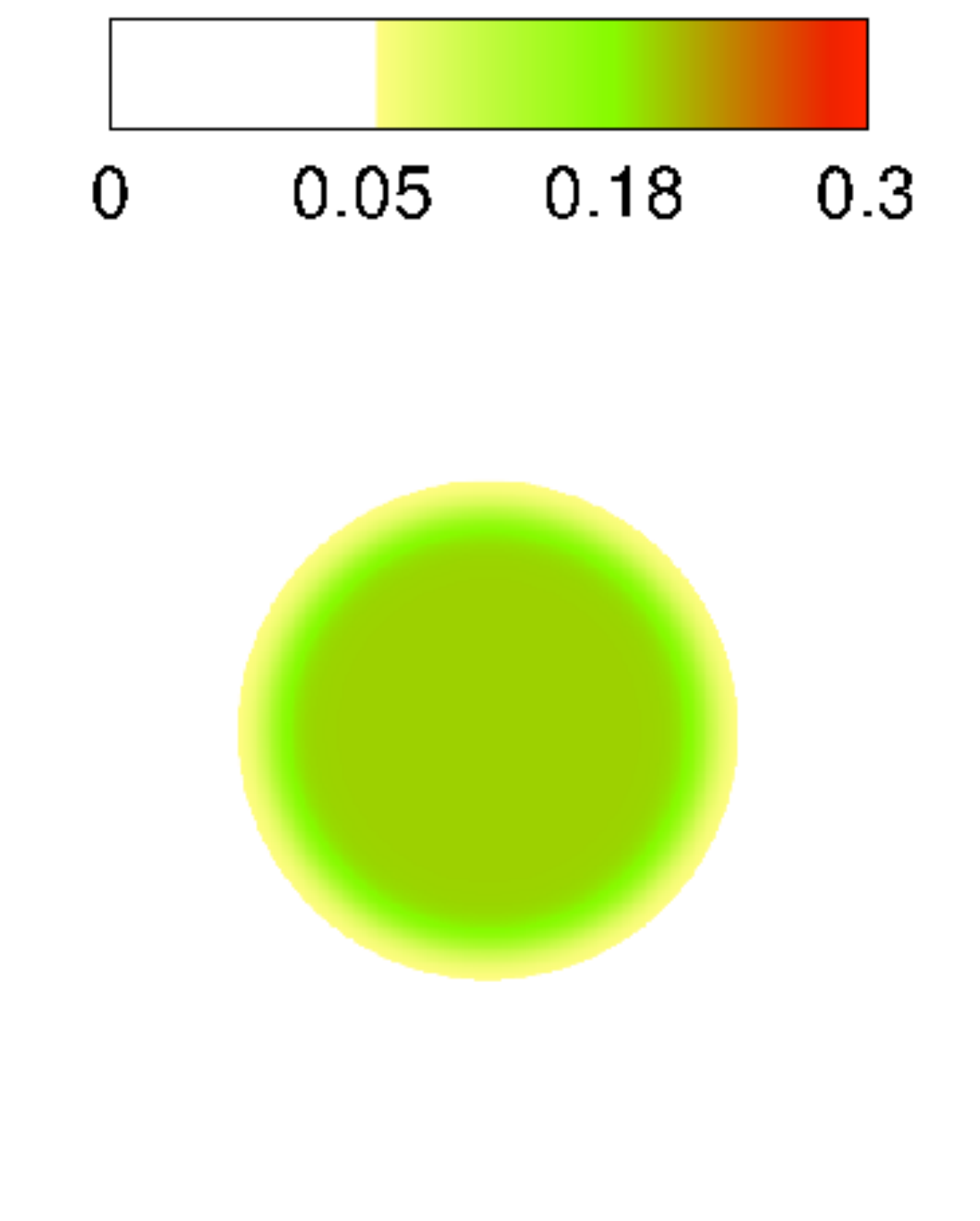}}
\subfigure[$t=10000$]{\includegraphics[width=1.5in]{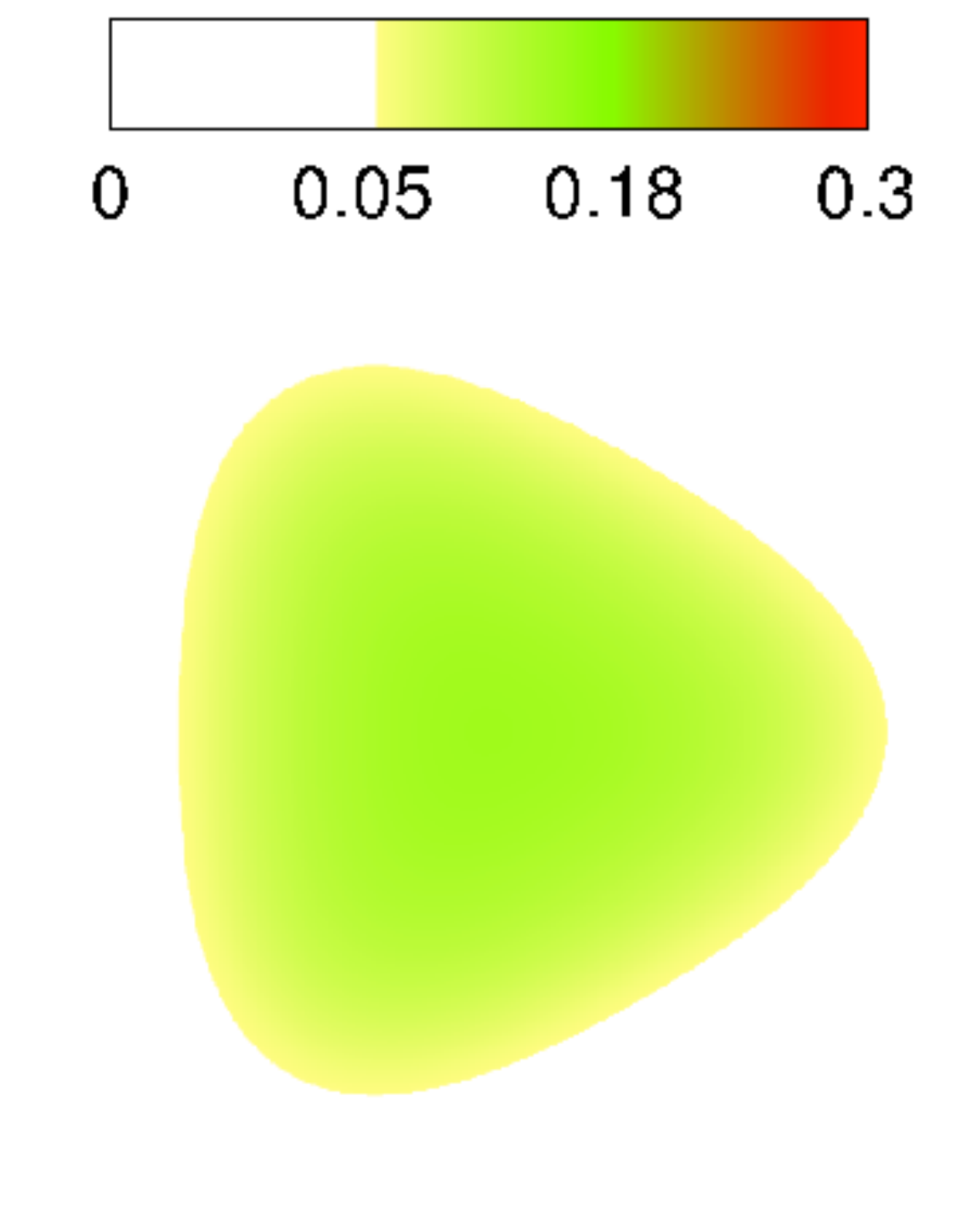}}
\caption{(Colour online) 
Spreading NLC droplet for $\cN=0.2$ and $s = -1/2$. (a) The anchoring condition 
at the substrate, $\phi(x,y)$, as defined in Eq.~(\ref{eq:numphi}). (b) The initial 
condition at $t=0$. (c) The droplet evolution at $t=10000$.
\mylab{fig:n02qnh}
}
\end{figure}

Figure~\ref{fig:n1qn1} shows the evolution of a spreading nematic droplet, for 
$\cN=1$. The anchoring at the substrate mimics the four-fold symmetric director 
structure near a defect of type $s=-1$, shown in Fig.~\ref{fig:n1qn1}(a). Again, 
due to the non-uniformity of this pattern in the radial direction, the droplet spreads 
in the manner that might be expected for the prescribed anchoring pattern. In 
addition, we observe rich pattern formation on the droplet surface, as illustrated by 
Fig.~\ref{fig:n1qn1} (b - c). 

\begin{figure}
\centering
\subfigure[anchoring condition]{\includegraphics[width=2.0in]{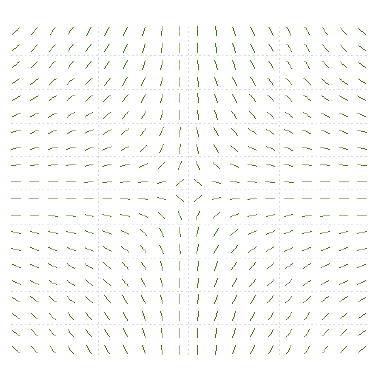}}
\subfigure[$t=500$]{\includegraphics[width=1.5in]{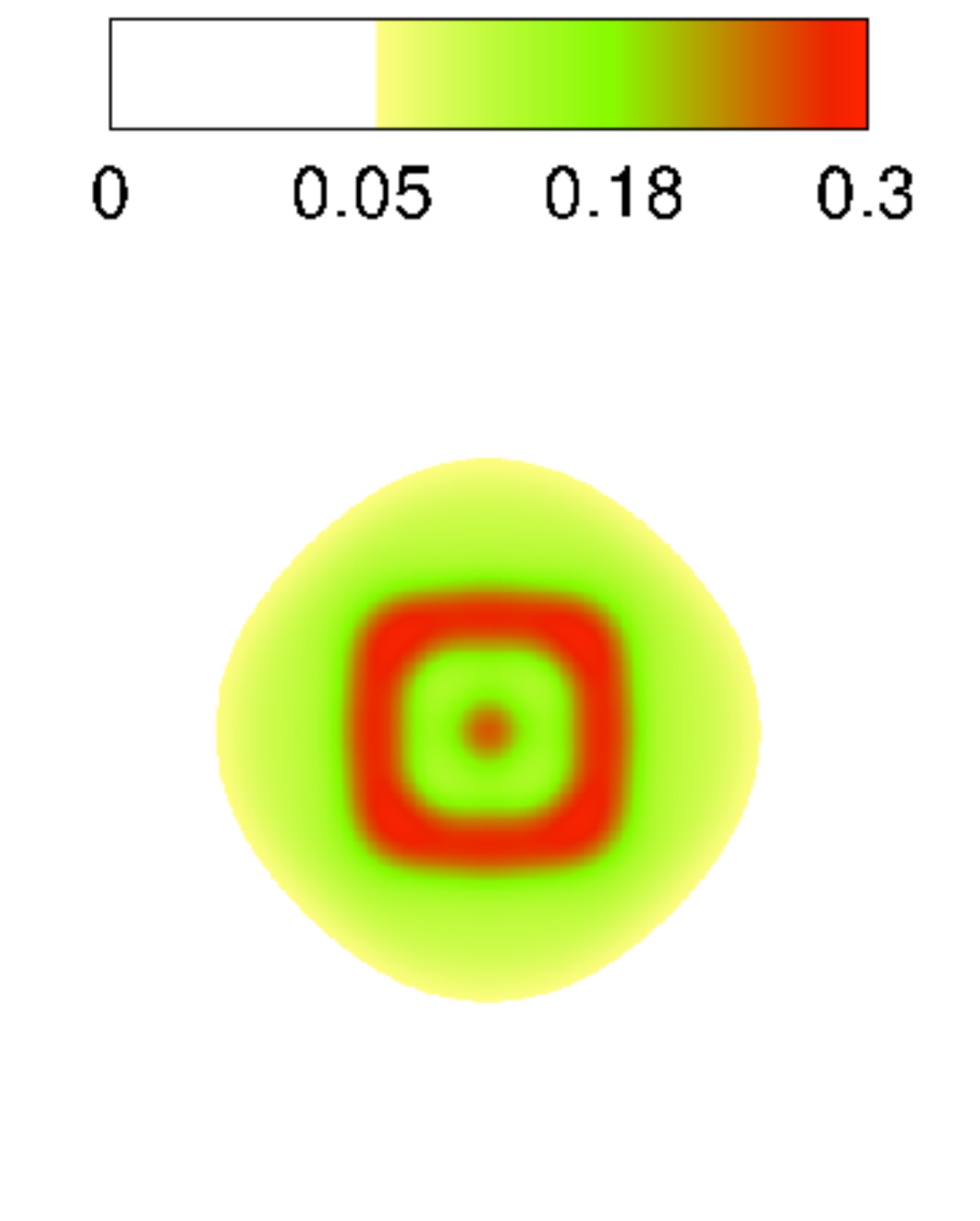}}
\subfigure[$t=1000$]{\includegraphics[width=1.5in]{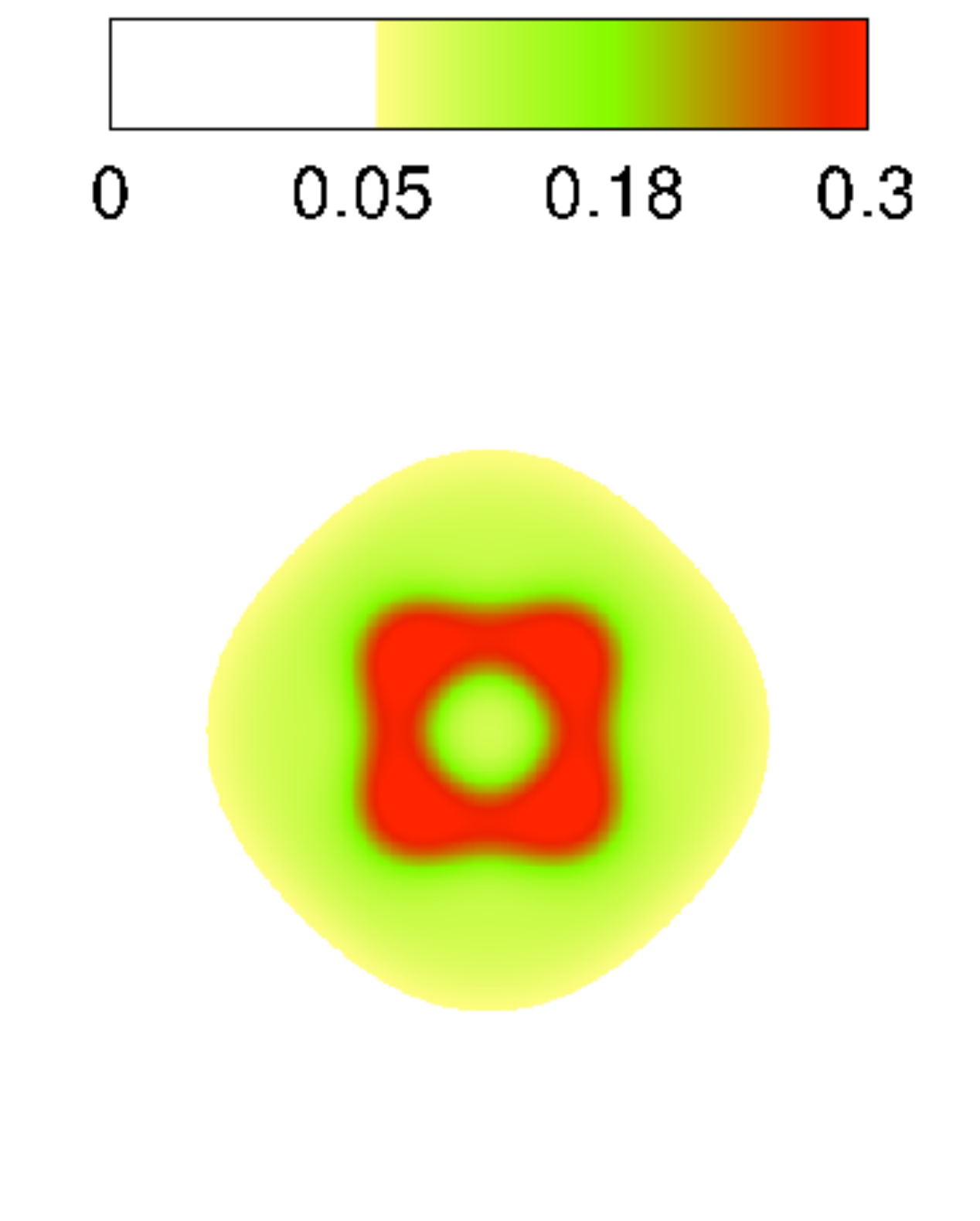}}
\caption{(Colour online) 
Spreading NLC droplet for $\cN=1$ and $s = -1$. The initial condition is the same 
as in Fig.~\ref{fig:n02qnh}(b). The anchoring condition at the substrate is shown in 
(a). The droplet evolution at $t=500$ and $t=1000$ are shown in (b) and (c), 
respectively.
\mylab{fig:n1qn1}
}
\end{figure}

By the analysis of \S~\ref{sec:LSA}, a flat film of the same thickness $h_0 = 0.2$ 
is unstable 
for $\cN=1$. The most unstable wavelength for this case is $l_m\approx 2\pi$, 
predicting that there will be about $3$ humps on the droplet surface for the 
chosen initial condition. We find that the results of simulations are consistent with 
these predictions, see, e.g., Fig.~\ref{fig:n1qn1}(b). The cross section in the radial 
direction of this figure shows $3$ hump-like structures, specified by (in 3D) one 
raised ring with one spherical hump at the centre. 

In order to confirm that this comparison between the LSA and nonlinear 
simulations extends to other parameter values, we next consider $\cN=10$. 
Here, the LSA predicts that the most unstable wavelength is $l_m\approx 1.6$ and 
the number of humps for the drop considered should be about $12$. 
Figure~\ref{fig:n1qh} shows this case, for a droplet spreading on an anchoring pattern
given by Eq.~(\ref{eq:numphi}) with $s=1/2$.  We again find remarkably good 
agreement between the LSA prediction for the flat film, and the observed simulation
for the cylindrical droplet. Figure~\ref{fig:n1qh}(b), for example, shows 
that there are $6$ rings that form, corresponding in the cross section to $12$ 
humps. Also note that the time scale for instability development is much shorter 
for $\cN=10$. Here, the unstable pattern has developed already at $t=5$, 
see Fig.~\ref{fig:n1qh}(b), while for  $\cN=1$ we have to wait 
until  $t=500$, as shown in Fig.~\ref{fig:n1qn1}(b). This finding is also 
consistent with the LSA predictions.  We remark also that the type of structures
seen in Fig.~\ref{fig:n1qh}(c) are reminiscent of certain free surface structures seen
in the experiments of~\citet{Poulard2005} (albeit within a much more complicated
setting in those experiments).

\begin{figure}
\centering
\subfigure[anchoring condition]{\includegraphics[width=2.0in]{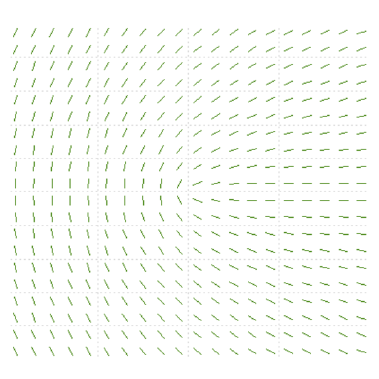}}
\subfigure[$t=5$]{\includegraphics[width=1.5in]{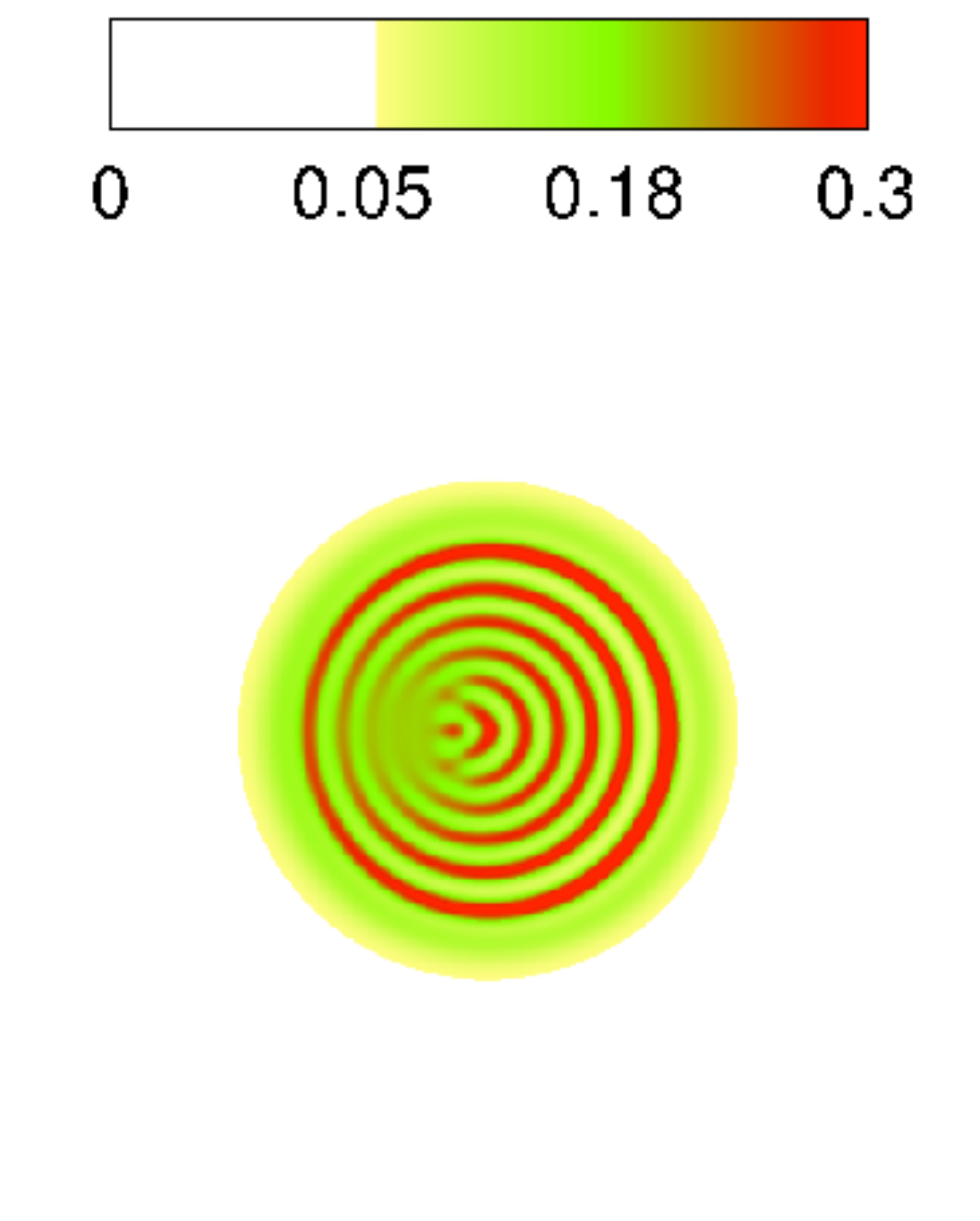}}
\subfigure[$t=30$]{\includegraphics[width=1.5in]{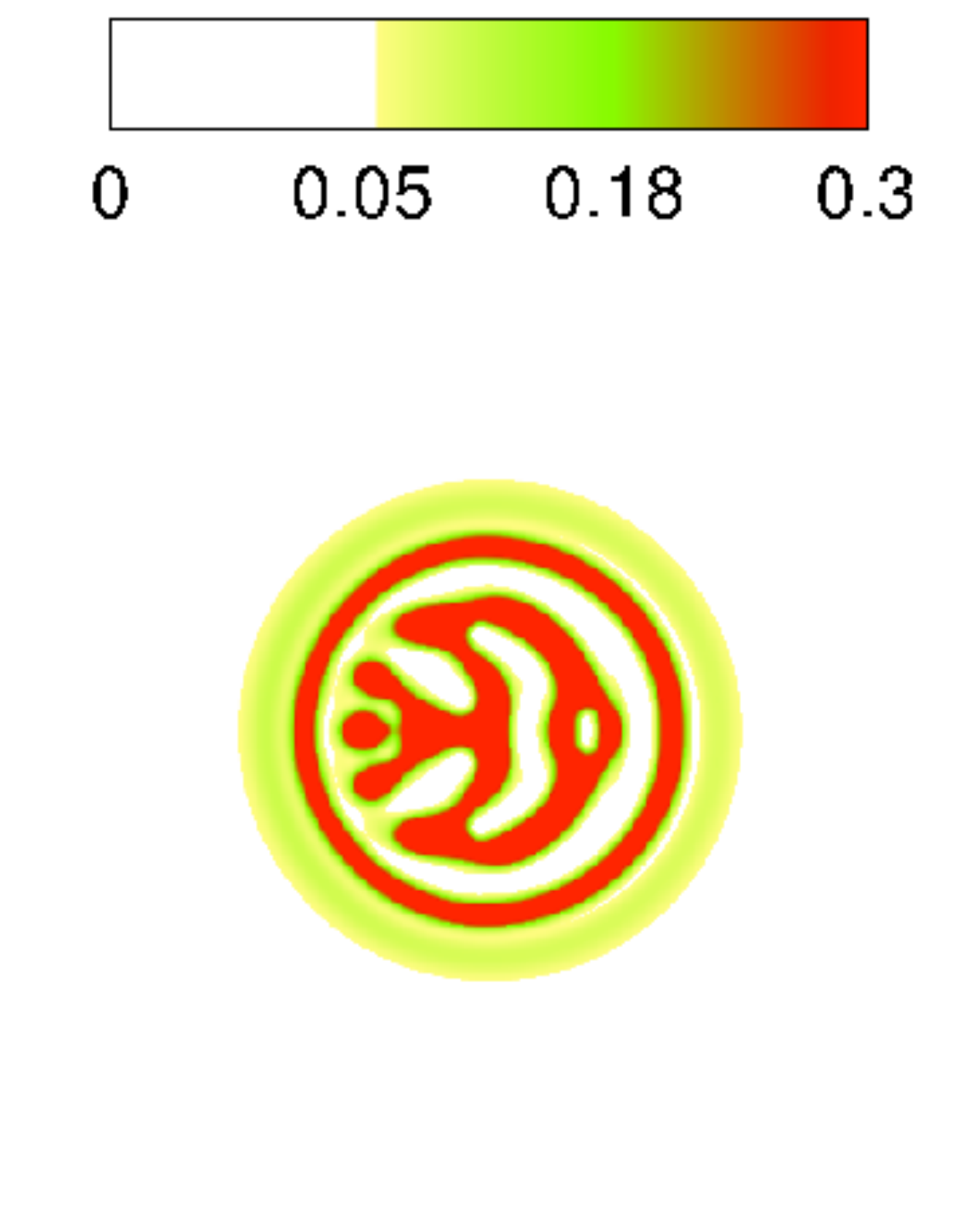}}
\caption{(Colour online) 
Spreading NLC droplet for $\cN=10$ and $s = 1/2$. The initial condition is the 
same as in Fig.~\ref{fig:n02qnh}(b). The anchoring condition at the substrate is 
shown in (a). The droplet evolution at $t=30$ is shown in (b).
\mylab{fig:n1qh}
}
\end{figure}

Finally, in Fig.~\ref{fig:n1q1} we show a spreading NLC droplet on a radially 
symmetric anchoring pattern, which mimics the director structure near a defect 
of type $s=1$.  The parameter $\cN$ is set to unity. Instead of using a cylindrical 
cap as the initial condition, here we choose one that is close to a spherical cap, 
shown as the solid (black) curve in Fig.~\ref{fig:n1q1}(b), a configuration for 
which our linear stability analysis is certainly not applicable. Nonetheless we 
note that, taking $h_0$ in the LSA to be the mean initial droplet height, one 
might anticipate instability for these parameter values. As the 
droplet spreads, the front remains circular for all time, while the surface exhibits 
radially symmetric instabilities, as anticipated. The instabilities 
appear as a ring (two humps in the cross section), which eventually closes up into
a single central hump for long times.

\begin{figure}
\centering
\subfigure[anchoring condition]{\includegraphics[width=2.0in]{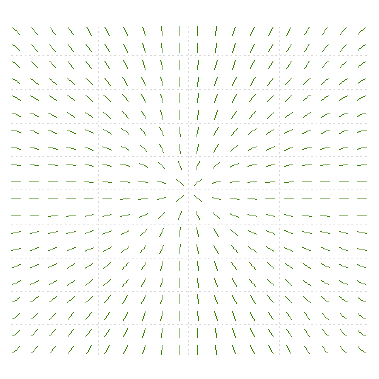}}
\subfigure[cross section]{\includegraphics[width=2.9in]{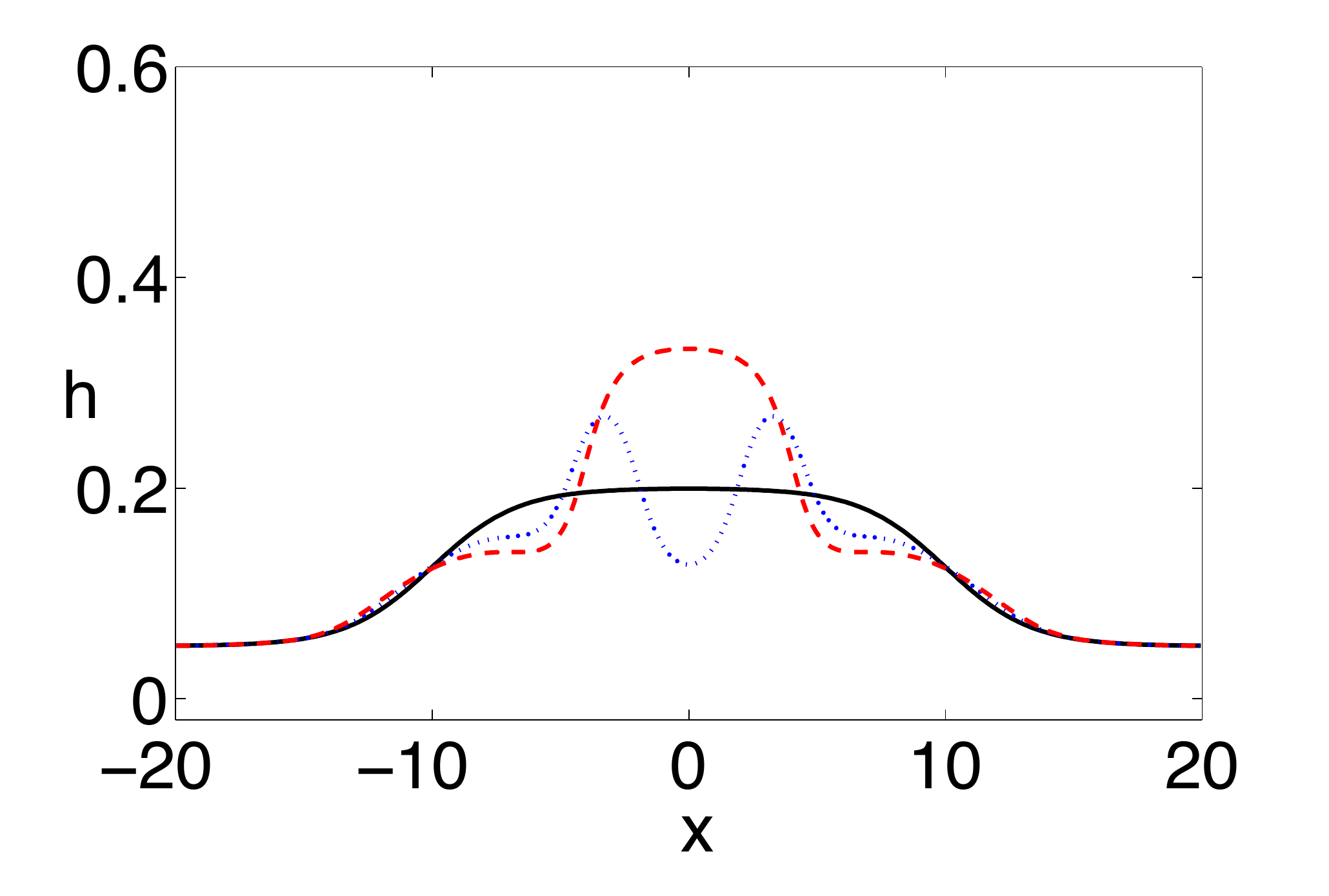}}
\subfigure[$t=0$]{\includegraphics[width=1.5in]{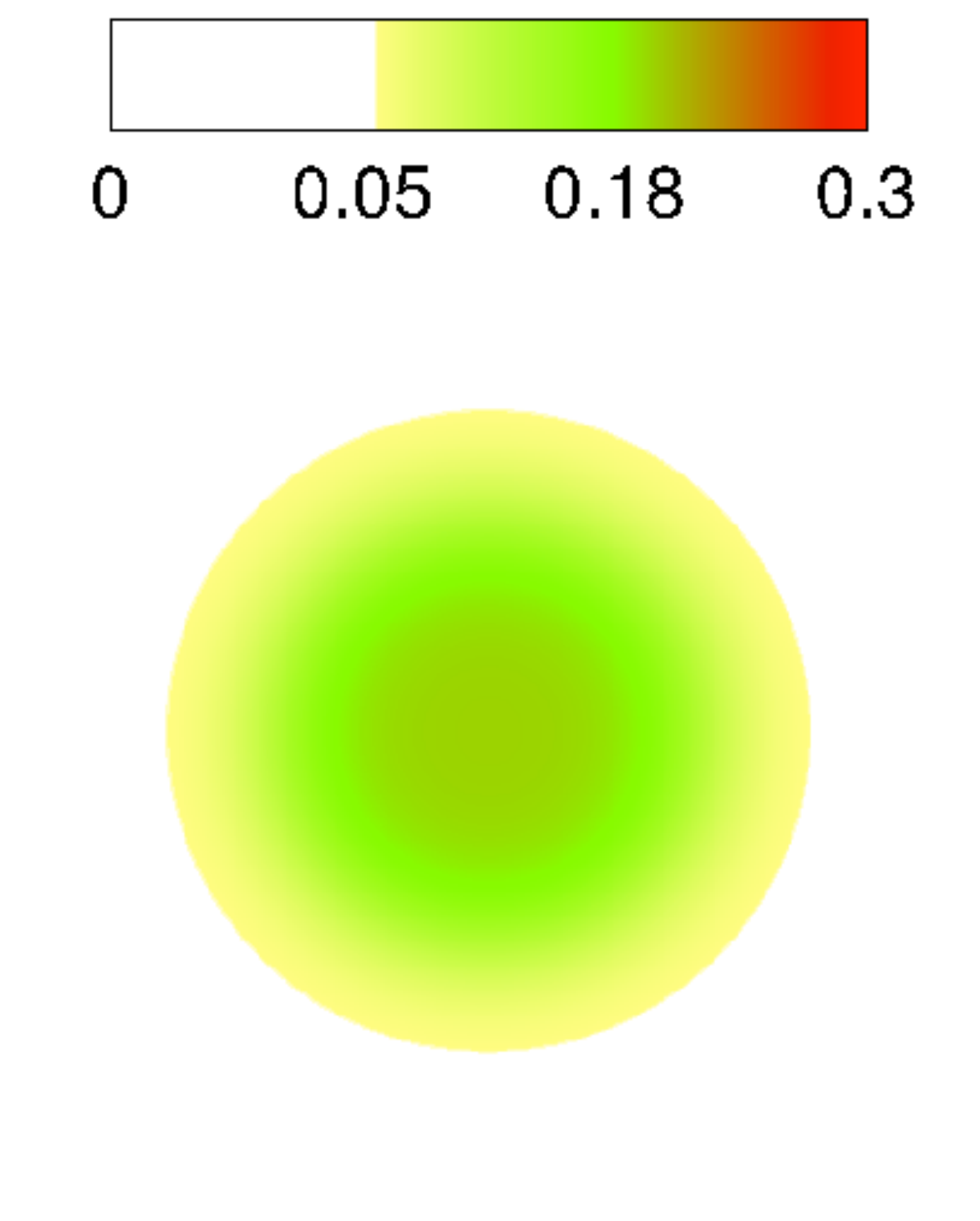}}
\subfigure[$t=300$]{\includegraphics[width=1.5in]{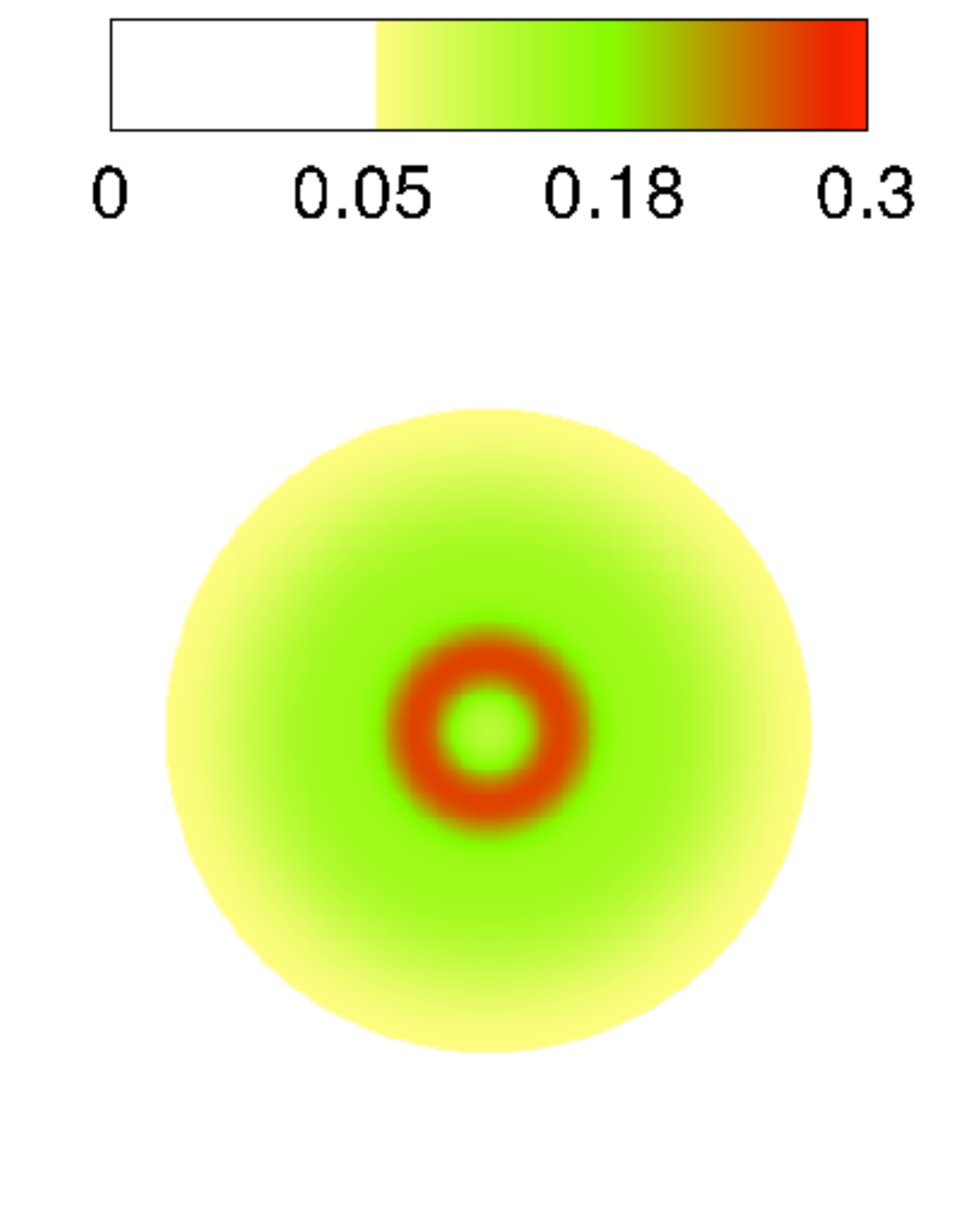}}
\subfigure[$t=600$]{\includegraphics[width=1.5in]{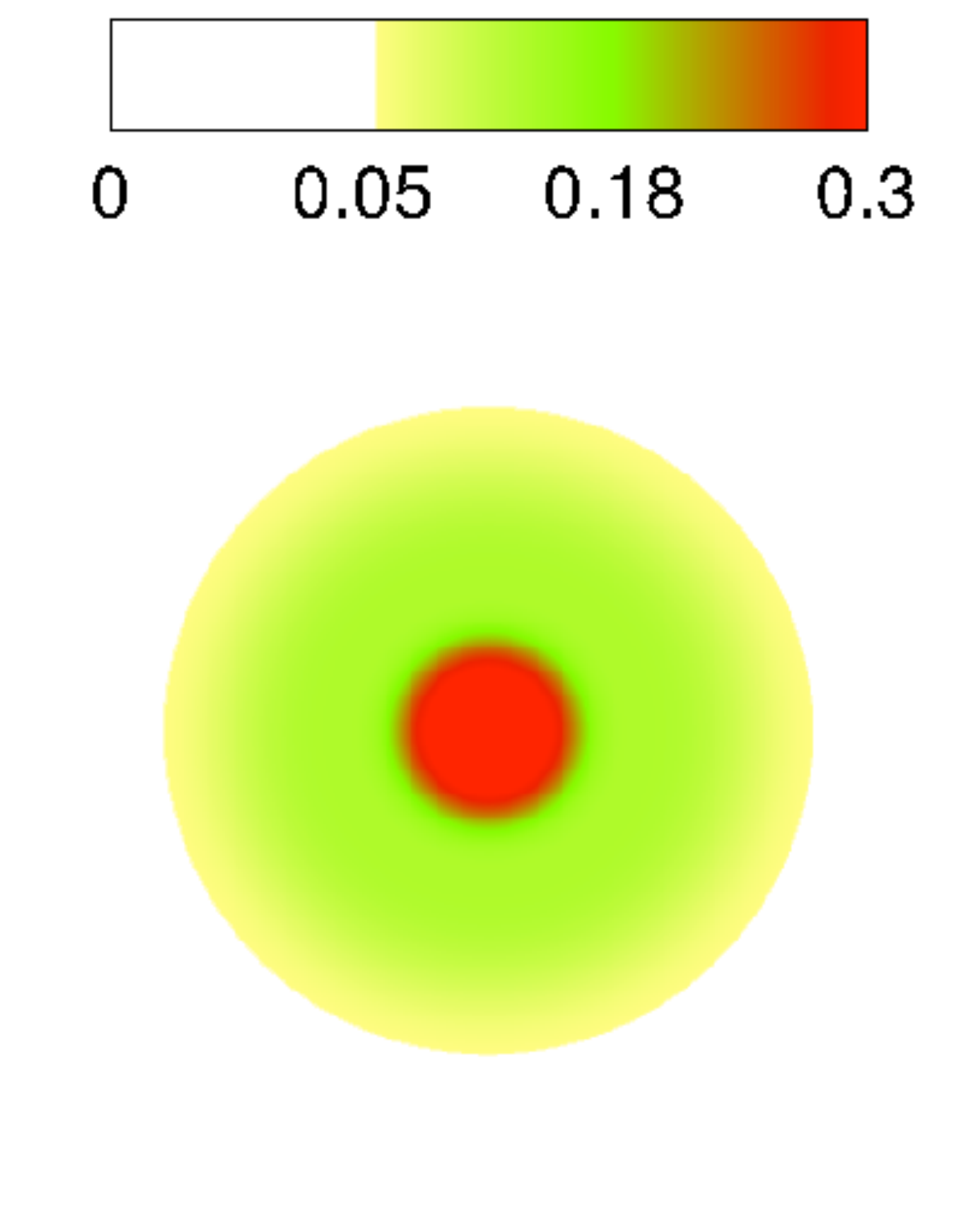}}
\caption{(Colour online) 
Spreading NLC droplet for $\cN=1$ and $s = 1$. The anchoring condition at the 
substrate is shown in (a). The cross sections of the droplet, $h(x, y=0, t)$, are 
shown in (b) for $t=0$ (solid (black) curve), $t=300$ (dotted (blue) curve) and 
$t=600$ (dashed (red) curve). The contour of the droplet at $t=0$, $t=300$ and 
$t=600$ are shown in (c), (d) and (e), respectively.
\mylab{fig:n1q1}
}
\end{figure}

\section{Conclusions}

We have presented a new model that describes three-dimensional spreading of 
thin films and droplets of nematic liquid crystal. To the best of our knowledge this 
is the first model of this kind to account for the effect of director variation in three 
dimensions on the shape of the overlying free surface. The stripe pattern of 
nematic films in a 3D setting was already analysed by \citet{Sparavigna1994}, 
\citet{Lavrentovich1994}, and \citet{Manyuhina2013} assuming the film remains a 
flat film. There the predicted instability mechanisms depend on the ratio of the 
various elastic constants while here the presented mechanism results from the 
coupling of free surface modulations and director orientation as described by the 
one-constant approximation.

Strong anchoring boundary conditions on the director at both boundaries are not 
suitable to describe a very thin spreading film (the director polar angle $\theta$ 
becomes singular at a contact line, leading to a strong diffusion, which is always 
stabilising). Instead, we impose 
weak conical surface anchoring on the polar angle $\theta$, with the anchoring 
energy given by Eq.~(\ref{eq:surface3}).  The anchoring at the substrate $z=0$ is 
taken to be strong and planar, with the azimuthal director angle $\phi (x,y,0)$ 
specified.  Our formulation preserves the property of strong anchoring when 
the film is thick, while allowing the director to relax to a state of planar alignment 
(though with anisotropy entering through nonuniform azimuthal patterning) when 
the film is very thin. The resulting equation for the film or droplet evolution is a 
fourth order nonlinear parabolic PDE, Eq.~(\ref{eq:NLC1}).

A simple linear stability analysis of Eq.~(\ref{eq:NLC1}) in the case of purely 2D 
flow predicts that a flat film may be unstable under certain conditions. The strong 
anchoring limit leads to a purely diffusive contribution from the elastic effects
that always acts stabilising; but weak anchoring can lead to instability. The 
physical mechanism is based on a coupling of the degrees of freedom of director 
orientation within the film and at its surfaces and the shape of the free surface 
itself. Even in the limit of instantaneous director relaxation considered here this 
coupling gives rise to an instability mechanism active in the film thickness range 
where anchoring and bulk elastic energies compete. A second mechanism 
that can lead to patterned spreading (which might be viewed as an instability in 
the advancing front) is that the anchoring condition on the azimuthal angle ($\phi$) 
at the solid substrate affects the speed of spreading. A drop spreads faster/slower 
when the substrate anchoring is parallel/perpendicular to the flow. For a substrate 
characterised by non-uniform anchoring conditions, the fluid front advances 
non-uniformly, in line with the prescribed anchoring patterns. This behaviour is 
exemplified by the analysis of a film spreading over a substrate with a striped 
anchoring pattern, leading to evolution with a sawtooth pattern in the advancing 
front, shown in Fig.~\ref{fig:stripe}.

We carried out numerical simulations of 3D spreading droplets for a variety of 
substrate anchoring patterns, focussing particularly on patterns that mimic the 
director structure near topological defects. Our simulations (including more than 
are reproduced here) indicate that (i) the flat film stability analysis serves as a 
remarkably good indicator of the stability of more complex spreading droplets, 
provided that the initial ``droplet height'' is well characterised; and (ii) although 
substrate anchoring clearly affects  spreading speed and the shape of the 
spreading front, it does not appear to influence the global free surface stability 
of spreading droplets.

Though simplified, the proposed model and the reported simulations provide 
valuable insight into the dynamics of spreading nematic droplets and films as 
observed experimentally by \citet{Poulard2005}, \citet{Delabre2009a} 
and \citet{Manyuhina2010}. The model as given by Eq.(\ref{eq:NLC1}) is rather 
general, relying only on the validity of the lubrication scaling (which in turn relies 
only on the droplet aspect ratio); the strong anchoring condition at the substrate; 
and the two-point trapezium rule approximation for the integral expressions 
appearing in Eq.~(\ref{eq:NLCT1}). Note that we propose and use a particular 
reasonable form for 
the anchoring function $m(h)$ only where it is necessary to carry out simulations or to 
demonstrate possible (in)stability regions. Thus, whenever the anchoring function 
$m(h)$ is obtained experimentally (as an empirical function to be fitted) 
Eq.~(\ref{eq:NLC1}), is applicable, and its predictions for the stability of a suitably 
thin flat film should be valid. 

However, there is still much to be done in order to elicit the full story in all its 
complexity. The results presented here, in particular regarding the influence of 
substrate anchoring patterns, clearly represent only a small subset of the possible 
spreading behaviour. Only very simple spreading scenarios and anchoring 
conditions are studied here, and it would clearly be of interest to simulate droplets 
spreading over more complex substrate patterning; for example, droplets 
spreading over several model defects as might be relevant in physical experiments. 
Our suggestion that the proposed substrate anchoring patterns may be thought of 
as idealised representations of defects in physical flows may of course also be 
questioned: it is known that the continuum nematic description used here breaks 
down in a small (nanometers) region around any defect, so our model cannot give 
an accurate description within such a defect core. Nonetheless, our simulations 
give some useful insight as to the effect that patterned planar anchoring can have 
on droplet evolution; and the similarity of Fig.~\ref{fig:n1qh}(c) to parts of Fig.2(c) 
in~\citet{Poulard2005} is intriguing.

While qualitatively illuminating, some aspects of our model are undoubtedly overly 
simplistic: our contact line regularisation may not adequately model the true 
physics as the ultra-thin precursor film is approached and molecular effects such 
as van der Waals' interactions become important; and indeed it may well not 
capture the true behaviour of the anchoring conditions. The effect of finite surface 
anchoring energy at the substrate may also need to be taken into consideration: 
the relative anchoring strengths at two bounding surfaces were found to be 
important in the transition behaviour of the director field (albeit in the presence of an 
applied electric field) by~\cite{Barbero1983} (see also citing
works). 
In future work we plan to introduce improved models for these aspects of 
the problem.

This work was supported by the NSF under grants DMS-0908158 and 
DMS-1211713. LJC also acknowledges financial support from King Abdullah 
University of Science and Technology under award no. KUK-C1-013-04, in the 
form of a Visiting Fellowship to the Oxford Centre for Collaborative Applied 
Mathematics.

\bibliographystyle{jfm}
\bibliography{NLC_JFM}

\end{document}